\documentclass[aps,pop,preprint,showpacs,showkeys]{revtex4}
\usepackage{amsmath}
\usepackage{amssymb}
\usepackage{graphicx}
\usepackage[usenames, dvipsnames]{color}

\newcommand{\Dk}{D_\text{Krook}}
\newcommand{\Drad}{D_\text{rad}}
\newcommand{\Dhyp}{D_\perp}
\newcommand{\kxeq}{k_x^\text{eq}}

\begin{document}
	\author{A.E.~Fraser$^1$}
	\author{M.J.~Pueschel$^2$}
	\author{P.W.~Terry$^1$}
	\author{E.G.~Zweibel$^1$}
	\affiliation{$^1$University of Wisconsin-Madison, Madison, Wisconsin 53706, U.S.A.\\
		$^2$Institute for Fusion Studies, University of Texas at Austin, Austin, Texas 78712, U.S.A.}
	\title{Role of stable modes in driven shear-flow turbulence}
\begin{abstract}
	A linearly unstable, sinusoidal $E \times B$ shear flow is examined in the gyrokinetic framework in both the linear and nonlinear regimes. In the linear regime, it is shown that the eigenmode spectrum is nearly identical to hydrodynamic shear flows, with a conjugate stable mode found at every unstable wavenumber. In the nonlinear regime, \textcolor{black}{turbulent} saturation of the instability is examined with and without the inclusion of a driving term that prevents \textcolor{black}{nonlinear} flattening of the mean flow, and a scale-independent radiative damping term that suppresses the excitation of conjugate stable modes. A simple fluid model for how momentum transport and \textcolor{black}{partial} flattening of the mean flow scale with the driving term is constructed, from which it is shown that\textcolor{black}{, except at high radiative damping,} stable modes play an important role in the turbulent state and yield significant\textcolor{black}{ly improved quantitative predictions when compared with corresponding models neglecting stable modes.}
\end{abstract}
\maketitle

\section{Introduction}
The prevalence of sheared flows in \textcolor{black}{diverse} systems has motivated their study for over a century. Their potential to drive instabilities and turbulence in fluids and plasmas is \textcolor{black}{central to angular momentum transport in astrophysical disks \cite{Balbus,Vanon}, to transport in the Earth's magnetosphere \cite{Faganello}, and to the possible generation and saturation of confinement-modifying zonal flows in fusion devices \cite{Rogers2000}}. The \textcolor{black}{linear} stability of simple shear flow configurations \textcolor{black}{has} been \textcolor{black}{thoroughly investigated from linear equations} \cite{Chandrasekhar,DrazinReid} and provides a rough understanding of the nature of more complex flow profiles early in their development, before unstable perturbations grow too large \cite{Gaster,Palotti}. However, as these flows develop beyond the regime of validity of linearized models, and nonlinear interactions between different components of the fluctuation become important, capturing \textcolor{black}{or understanding} their behavior with \textcolor{black}{any set of constructs based on linear analysis becomes problematic} \cite{Liou,Nikitopoulos,Horton}.

Instead, studies generally \textcolor{black}{rely on} direct numerical simulations to investigate relevant physical effects \cite{Faganello,Palotti,Henri}. \textcolor{black}{In many cases of interest, these methods cannot produce solutions for} physically relevant parameters, such as the high Reynolds numbers found in astrophysical systems\textcolor{black}{. This motivates} the development of scaling models that can inform how the system \textcolor{black}{extrapolates} to parameter regimes inaccessible to simulations. \textcolor{black}{Valid scaling models require an understanding of the physics of all relevant phenomena, including turbulent responses that modify the unstable flow, like nonlinear fluctuation structures, cascades, and momentum transport.}

\textcolor{black}{Regarding} nonlinear \textcolor{black}{processes} that become relevant as the linear growth phase ends, recent analytical work \textcolor{black}{on shear-flow instability saturation} has demonstrated \textcolor{black}{the importance of fluctuation dissipation that arises at large scales due to excitation of stable modes} \cite{Paper1}. When an unstable shear flow is perturbed from equilibrium, these \textcolor{black}{linear} modes \textcolor{black}{are generally a part of the initial perturbation, decaying from their small initial amplitude. Given this initial decay, stable modes} are \textcolor{black}{typically} ignored \textcolor{black}{in} constructing reduced nonlinear models that \textcolor{black}{draw from linear physics} \cite{Liou,Nikitopoulos,Horton}. However, nonlinear interactions with unstable modes \textcolor{black}{can} drive \textcolor{black}{stable} modes to large amplitude. \textcolor{black}{Because they are linearly stable}, they \textcolor{black}{provide} a route for energy to \textcolor{black}{be removed from fluctuations} at large scales, before it is able to cascade to small scales\textcolor{black}{, thereby modifying the flow, its spectrum, and its transport \cite{Paper1}.} This \textcolor{black}{represents} a significant \textcolor{black}{departure from} the usual picture of instability-driven turbulence, where energy injection by unstable modes is assumed to be balanced by \textcolor{black}{conservative} nonlinear energy transfer to small, dissipative scales.

While it has been shown that nonlinear interactions with large-scale stable modes \textcolor{black}{can be} important in saturating shear-flow instabilities, their amplitude and contribution to the fluctuating flow and momentum transport in fully-developed turbulence remains an open question, which we pursue in this paper. Additionally, we explore whether reduced models of shear-flow-driven turbulence that are based \textcolor{black}{solely on the} linear instability might be improved by including the effects of large-scale stable modes. This is a natural expectation given their importance in saturating the instability, their introduction of a large-scale linear energy sink, and their potential to modify momentum transport. This is also motivated by recent work in the context of instability-driven turbulence in fusion devices, where reduced turbulence models that include details of stable modes and instability saturation physics have shown to be effective \cite{Terry2018,Hegna,Whelan}.

We address these questions by performing direct numerical simulations of an unstable shear flow that develops into turbulence, and comparing the contribution of different linear modes to the turbulent flow and the Reynolds stress. Our simulations are performed using the gyrokinetic \textcolor{black}{turbulence} code \textsc{Gene} \cite{genecode,Jenko}, which has previously been used to examine stable modes in other turbulent systems \cite{HatchLeft,TerryLeft}, and \textcolor{black}{includes both initial value and eigenvalue solvers}. This allows us to benchmark our calculated growth rates against previous gyrokinetic studies of the same system \cite{Rogers2005}, as well as investigate differences between shear flow instabilities in hydrodynamics and gyrokinetics with regards to both the linear mode spectrum and instability saturation. In particular, while it is understood that all unstable, inviscid, incompressible, two-dimensional (2D) hydrodynamic flows include one stable mode for every unstable mode \cite{DrazinReid}, and previous work has shown that these stable mode are nonlinearly driven in the fluid system \cite{Paper1}, whether these results apply to the gyrokinetic case as well has not been explored. 
To allow for more direct comparisons with previous work, all simulations presented in this paper are effectively 2D, with no variations in the direction of the strong guide field ($k_z = 0$).

The flow we examine is a sinusoidally-varying $E \times B$ parallel shear flow with periodic boundary conditions. \textcolor{black}{The hydrodynamic counterpart to this flow is often referred to as Kolmogorov flow when it is maintained by a constant forcing term \cite{Platt,Musacchio,Lucas}.} This flow profile is particularly relevant to astrophysical disks, where its \textcolor{black}{Kelvin-Helmholtz (KH)} instability is studied as a saturation mechanism for the magnetorotational instability \cite{Goodman,PessahGoodman,Pessah,Latter2009,Latter2010,Longaretti} or its collisionless counterpart \cite{Squire}, and in fusion devices, where it is studied as a potential secondary and tertiary instability to streamers and zonal flows  \cite{Rogers2000,Kim}. In order to admit a quasi-stationary state of driven turbulence where energy dissipation is balanced by energy injection, we continually reinforce the mean flow using a Krook operator previously \textcolor{black}{employed} similarly to reinforce current gradients in tearing mode studies \cite{PueschelKrook}\textcolor{black}{, and referred to as a linear relaxation term in studies of barotropic jets \cite{MarstonKrook}}. With this forcing term, the system bears a strong resemblance to Kolmogorov flow \cite{Platt,Musacchio,Lucas}, \textcolor{black}{with the exception that it is not a constant forcing}. From a numerical perspective, Kolmogorov flow presents a convenient choice of unstable shear flow to study due to its simple description in a Fourier basis and the lack of no-slip boundary conditions that could otherwise generate boundary layers. This also allows us to address whether the saturation physics \textcolor{black}{active} in the free shear layer \cite{Paper1} \textcolor{black}{is applicable} to a driven periodic shear flow.

Our simulations also include damping terms in the form of hyperdissipation and scale-independent radiative damping. The form of the radiative damping term is such that it damps every mode equally. In systems with pairs of stable and unstable modes, this disproportionately affects the stable mode \textcolor{black}{amplitude} relative to the unstable one \textcolor{black}{in the nonlinear state} \cite{Terry2009}. Thus, varying the degree of radiative damping in our system allows us to assess whether different shear-driven turbulence regimes exist with significantly different stable mode effects, and how these regimes might differ.

The remainder of this paper is organized as follows. Section II starts with a brief review of hydrodynamic parallel shear flows for comparison with our gyrokinetic results, as well as some unique aspects of the particular flow profile studied here, followed by a discussion of the numerical implementation used in our work, including the specific forms of forcing and dissipation. In Sec.~III we show the full eigenmode spectrum for the gyrokinetic KH instability. A description of the nonlinear evolution of the flow is presented in Sec.~IV, where we discuss saturation and decaying turbulence when forcing is absent, driven turbulence with external forcing, and turbulent momentum transport in this system. Section V examines the turbulence in terms of the role played by the linear eigenmodes, and compares reduced descriptions and scaling models of the turbulence with and without stable modes. Conclusions are presented in Sec.~VI.

Throughout this paper, we adopt the notation that $\hat{A}(x, k_y)$ denotes the Fourier transform in $y$ of $A(x, y)$, and $\tilde{A}(k_x, k_y)$ denotes the Fourier transform in $x$ and $y$.

\section{Shear Flow Instability}

\subsection{Rayleigh's Stability Equation}
The stability of parallel shear flows is generally investigated by examining infinitesimal perturbations about equilibrium solutions to the Navier-Stokes equation. When considering a 2D, inviscid, incompressible flow that is perturbed from an equilibrium, the vorticity equation becomes
\begin{equation} \label{nonlinear vorticity}
\frac{\partial}{\partial t}\nabla^2\phi + V\frac{\partial}{\partial y}\nabla^2\phi - \frac{d^2 V}{dx^2}\frac{\partial \phi}{\partial y} + \frac{\partial \phi}{\partial x} \frac{\partial}{\partial y} \nabla^2\phi - \frac{\partial \phi}{\partial y}\frac{\partial}{\partial x}\nabla^2\phi = 0,
\end{equation}
where $V(x)$ is the $y$-directed equilibrium shear flow, and $\phi(x,y,t)$ is the streamfunction of the perturbation $\mathbf{v} = \nabla \phi \times \hat{\mathbf{z}}$. 
The linear dynamics can then be explored by dropping the nonlinearities and using the normal mode ansatz 
\begin{equation}\label{Fourier expansion}
\phi(x,y,t) = \sum_{k_y} \sum_j \hat{\phi}_j(x,k_y)e^{i(k_y y + \omega_j t)},
\end{equation}
yielding
\begin{equation}\label{Rayleigh's}
(\omega_j + k_y V)\left(\frac{\partial^2}{\partial x^2}-k_y^2\right)\hat{\phi}_j - k_y \hat{\phi}_j \frac{d^2 V}{dx^2} = 0.
\end{equation}
Equation \eqref{Rayleigh's} is known as Rayleigh's stability equation, or as the Orr-Sommerfeld equation when the effect of viscosity on $\phi$ is included. It can be \textcolor{black}{solved} as an eigenvalue problem, \textcolor{black}{yielding} a set of eigenvalues $\omega_j$ and eigenmodes $\hat{\phi}_j$, with $j$ enumerating \textcolor{black}{the} eigenmode\textcolor{black}{s} at a given $k_y$. \textcolor{black}{The eigenvalue} $\omega_j$ is complex, with real frequency $\mathrm{Re}(\omega_j)$ and growth rate $\gamma_j = -\mathrm{Im}(\omega_j)$. If any eigenmode has a positive growth rate, the flow is unstable. Furthermore, taking the complex conjugate of Eq.~\eqref{Rayleigh's} shows that for each unstable solution there exists a stable solution with equal and opposite growth rate \cite{DrazinReid}. Previous work \cite{Paper1} demonstrated that nonlinear interactions with these stable modes play an important role in saturating the growth of unstable modes. In \textcolor{black}{the present} work we perform nonlinear simulations of an unstable shear flow and examine the role played by stable modes beyond the onset of saturation.

\subsection{Kolmogorov Flow}
One unstable flow profile of relevance in fusion and astrophysical systems is a sinusoidal equilibrium flow with periodic boundary conditions \cite{Goodman,PessahGoodman,Pessah,Latter2009,Latter2010,Longaretti,Squire,Vanon,Rogers2000,Rogers2005}.
For a sinusoidal flow profile $V(x) = V_0 \cos(\kxeq x)$ in a periodic domain, Eq.~\eqref{Rayleigh's} lends itself well to being solved using spectral methods. Defining $\tilde{\phi}_j(k_x, k_y)$ as the Fourier series expansion of $\hat{\phi}_j(x, k_y)$, the Fourier representation of Eq.~\eqref{Rayleigh's} is
\begin{equation}\label{Fourier Rayleigh's}
\omega_j(k_x^2+k_y^2)\tilde{\phi}_j + \frac{k_y V_0}{2}\left[(k_x^2-2k_x\kxeq+k_y^2)\tilde{\phi}_j^{-}+(k_x^2+2k_x\kxeq+k_y^2)\tilde{\phi}_j^{+}\right]=0,
\end{equation}
where $\tilde{\phi}_j^\pm \equiv \tilde{\phi}_j(k_x \pm \kxeq, k_y)$. Equation \eqref{Fourier Rayleigh's} immediately demonstrates that each eigenmode exhibits a \textcolor{black}{discrete,} comb-like structure when viewed through a Fourier transform\textcolor{black}{: f}or a given eigenmode $\hat{\phi}_j(x, k_y)$, if its Fourier transform $\tilde{\phi}_j(k_x, k_y)$ is nonzero at some $k_x$, then it is also nonzero at $k_x + n\kxeq$ for every integer $n$ \textcolor{black}{(though $\tilde{\phi}_j$ is still expected to fall off at large $|k_x|$, so that calculations with a finite number of $k_x$ can be expected to capture the structure well)}. This property of the system will have important consequences in later sections when we compare simulations with different box sizes, and when we explore the possibility of approximating the turbulent state by truncating the summation over $j$ in Eq.~\eqref{Fourier expansion} to a reduced number of modes.

\subsection{Numerical implementation and benchmarking}
We perform simulations of a KH-unstable sinusoidal $E \times B$ flow using the gyrokinetic framework \cite{Brizard} as implemented in the \textsc{Gene} code \cite{genecode,Jenko}. \textcolor{black}{The gyrokinetic framework applies to systems with a strong guide field, where the parallel length scale of fluctuations is much larger than the perpendicular length scale, and the relevant frequencies are much smaller than the ion cyclotron frequency.} The use of gyrokinetics \textcolor{black}{for this work} is motivated by \textsc{Gene}'s unique tools for performing eigenmode decompositions \cite{HatchLeft,TerryLeft}. We simulate a system with two spatial dimensions, with a $y$-directed flow that varies sinusoidally in $x$, a strong guide field in the $z$ direction, and \textcolor{black}{no variations} in $z$. Our simulation domain is a periodic box of dimensions $L_x \times L_y$ with no curvature or magnetic shear. The flow arises \textcolor{black}{from} the $E \times B$ drift of the particles, \textcolor{black}{allowing} the electrostatic potential $\phi$ \textcolor{black}{to} serve as the streamfunction for the flow. We model the plasma with gyrokinetic ions and electrons with hydrogen mass ratio, ion and electron background temperatures $T_\mathrm{i} = T_\mathrm{e}$, no collisions, and no electromagnetic fluctuations (plasma $\beta = 0$).

We drive instability with a potential and corresponding distribution function that vary sinusoidally in $x$. \textsc{Gene} uses a $\delta f$ formalism, where the full distribution function is separated into equilibrium $F_0$ and fluctuation $f$, with the code solving for the evolution of the fluctuation. We let $f(x, y, v_\parallel, \mu, s, t)$ and $\tilde{f}(k_x, k_y, v_\parallel, \mu, s, t)$ denote the (guiding-center) distribution function for species $s$ in real and Fourier space. For the remainder of this paper, we will generally use notation that suppresses the species and velocity dependence of $f$, and instead focus on its dependence on the spatial coordinates and time.

For benchmarking against previous work \cite{Rogers2005}, the instability is first examined by implementing the sinusoidal flow \textcolor{black}{with low-amplitude white noise} as an initial condition in the fluctuation\textcolor{black}{,} formally evolv\textcolor{black}{ing the system} nonlinearly, with a homogeneous Maxwellian equilibrium $F_0$. This corresponds to solving the equation
\begin{equation}\label{RogersEq}
\frac{\partial f}{\partial t} = \left\{ f, \bar{\phi}\right\}
\end{equation}
with a sinusoidal initial condition $f(t=0), \phi(t=0) \sim \sin(\kxeq x)$ and low-amplitude noise to seed instability. The only term on the right-hand side of Eq.~\eqref{RogersEq},
\begin{equation}\label{ExB NL def}
\left\{f, \bar{\phi}\right\} \equiv \frac{\partial f}{\partial x}\frac{\partial \bar{\phi}}{\partial y} - \frac{\partial \bar{\phi}}{\partial x}\frac{\partial f}{\partial y},
\end{equation}
is the $E \times B$ nonlinearity, whose Fourier transform becomes $\sum_{k_x', k_y'} \left( k_x'k_y - k_xk_y' \right) \tilde{\bar{\phi}}(k_x', k_y')\tilde{f}(k_x-k_x', k_y-k_y')$. Here, $\bar{\phi}$ is the gyro-averaged $\phi$, whose Fourier transform is given by $\tilde{\bar{\phi}}(k_x, k_y, \mu, s) = J_0(\sqrt{k_x^2 + k_y^2} \rho)\tilde{\phi}(k_x,k_y)$, where $J_0$ is a Bessel function, and $\rho$ is the gyroradius of species $s$ with magnetic moment $\mu$. The code evolves $f$ according to Eq.~\eqref{RogersEq} and calculates $\phi$ using Gauss's law as described in Refs.~\cite{Merz} and \cite{Pueschel2011}. \textcolor{black}{The normalizations used by \textsc{Gene} are described in Ref.~\cite{Merz}; however, in this paper we will follow the standard convention used in the fluids community and normalize quantities with respect to the equilibrium flow velocity $V_0$ and its wavelength $\kxeq$, which are normalized in the code by $V = V_\text{phys}\rho_\text{ref}c_\text{ref}/L_\text{ref}$ and $k_x = k_{x \text{phys}} / \rho_\text{ref}$.}

\begin{figure}
	\includegraphics[width=16cm]{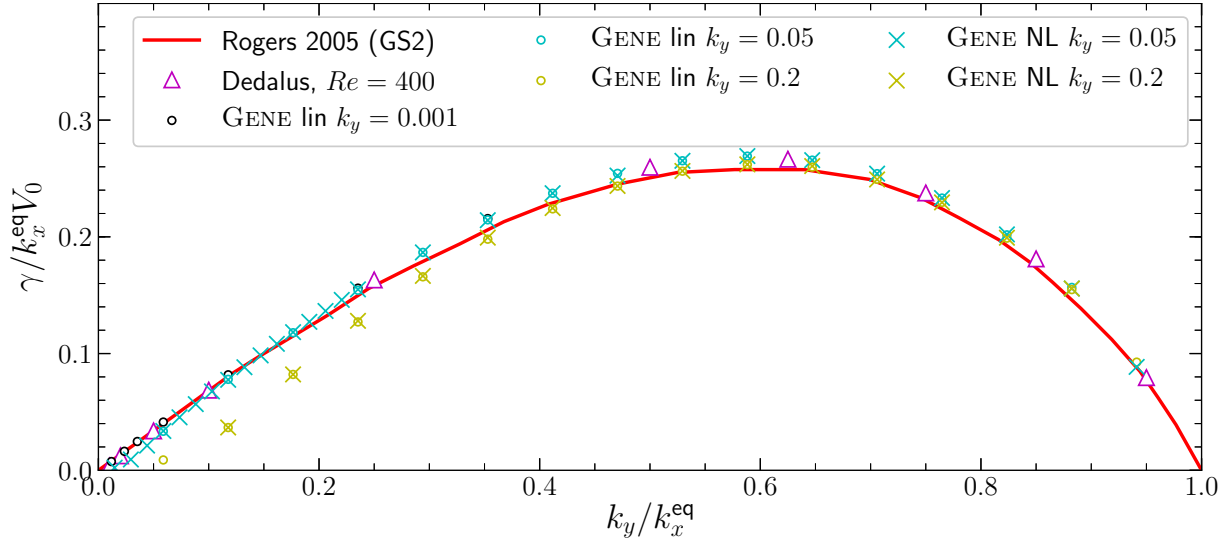}
	\caption{Dispersion relation for \textcolor{black}{the} KH instability of a sinusoidal flow $\mathbf{V} = V_0\cos(\kxeq x) \hat{\mathbf{y}}$. Growth rate $\gamma$ is plotted against the wavenumber $k_y$ of the perturbation, with $\gamma$ normalized to the equilibrium shear $\kxeq V_0$ and $k_y$ to the equilibrium wavenumber $\kxeq$. Crosses are obtained from nonlinearly evolving a perturbed sinusoidal flow in \textsc{Gene} according to Eq.~\eqref{RogersEq}, while dots are from solving the linear Eq.~\eqref{Linear GK eqn}. Results compare well with both previous gyrokinetic simulations (red curve\textcolor{black}{, see Ref.~\cite{Rogers2005}}) and hydrodynamic simulations of an equivalent system (\textcolor{black}{magenta} triangles). The stabilization of the $k_y=0.2$ points at low $k_y/\kxeq$ (i.e.~high $\kxeq$) can be attributed to finite Larmor radius effects. All modes have zero real frequency.}
	\label{dispersion}
\end{figure}

Consistent with fluid theory, our system is unstable to perturbations of the same form as Eq.~\eqref{Fourier expansion} for a range of perturbation wavenumbers $k_y$, with the growth rate scaling with the base flow amplitude $V_0$. \textcolor{black}{Growth rates} from this \textcolor{black}{formally} nonlinear setup are indicated by crosses in Fig.~\ref{dispersion}. For direct comparison with previous work \cite{Rogers2005}, the wavenumber of the equilibrium $\kxeq$ was varied at fixed $k_y$, where perturbations are unstable for $0 < k_y/\kxeq < 1$. For this reason, in the remainder of this paper we focus our discussion on modes that lie in this range.

As demonstrated in Fig.~\ref{dispersion}, nonlinear simulations with appropriate initial conditions can be used to investigate some of the linear dynamics of this system, such as the growth rate and mode structure of the most unstable mode at each $k_y$. However, to solve for other linear modes, which are known to exist in fluid models \cite{DrazinReid}, terms corresponding to interaction with the driving flow need to be implemented as a linear operator, so that Eq.~\eqref{RogersEq} can be linearized similar\textcolor{black}{ly} to Eq.~\eqref{Rayleigh's}, in the form
\begin{equation}\label{Linear GK eqn}
\frac{\partial f}{\partial t} = \mathcal{L}_\text{KH}[f]
\end{equation}
for a linear differential operator $\mathcal{L}_\text{KH}$. To that end, we have implemented the linear operator $\mathcal{L}_\text{KH}$ in the \textsc{Gene} code. This allows computations to be performed with $\mathcal{L}_\text{KH}[f]$ on the right-hand side of the equation for $\partial_t f$ for
\begin{equation}\label{LinearOp}
\mathcal{L}_\text{KH}[f] \equiv \left\{ f_0, \bar{\phi}\right\} + \left\{ f, \bar{\phi}_0 \right\},
\end{equation}
where $\phi_0$ is the \textcolor{black}{electrostatic potential (streamfunction) for the} sinusoidal base flow, and $f_0$ is the self-consistent distribution function \textcolor{black}{corresponding to} $\phi_0$. Specifically, we use
\begin{equation}
f_0(s) = \frac{V_0}{\kxeq}\frac{\delta_{k_x, \kxeq} - \delta_{k_x, -\kxeq}}{2i}F_0(s)\frac{q_s}{T_s}J_0\frac{1-\Gamma_0}{\Gamma_0},
\end{equation}
where $\delta_{k_x, k_x'}$ is the Kronecker delta, $F_0(s)$, $q_s$, and $T_s$ are the equilibrium Maxwellian, charge, and temperature of species $s$, and the Bessel functions $J_0$ and $\Gamma_0$ relate to finite Larmor radius (FLR) effects as detailed in Refs.~\cite{Merz} and \cite{Pueschel2011}. This form of $f_0$ is used for secondary instability tests in tokamak-relevant systems \cite{Pueschel2013a}, and yields a sinusoidal $\phi_0(x)$ corresponding to a sinusoidal equilibrium flow in the $y$ direction with amplitude $V_0$ and wavenumber $\kxeq$. 

Note that $\mathcal{L}_\text{KH}$ has $x$ dependence but no $y$ dependence, so its eigenmodes will have Fourier dependence in $y$ and more complex structure in $x$, similar to the hydrodynamic case discussed in Sec.~II.~A. The dots in Fig.~\ref{dispersion} are obtained by solving Eq.~\eqref{Linear GK eqn}, and their agreement with the corresponding crosses demonstrates successful implementation of the linear drive. For both setups, convergence checks were performed in spatial and velocity coordinates. Well-converged growth rates generally require $33$ grid points in $x$, though far fewer points were required for $k_y/\kxeq \lesssim 0.5$. For the remainder of this paper, results are presented with $V_0 = 10$ and $\kxeq = 0.1$ using the linearized $\mathcal{L}_\text{KH}$. A convenient consequence of these parameters is that times and frequencies have the same value when expressed in standard \textsc{Gene} normalizations as they do in typical normalizations used in calculations of unstable shear flow in the fluids community, where $t$ is often measured in units of $(\kxeq V_0)^{-1}$.

Cyan triangles in Fig.~\ref{dispersion} are obtained from solving the Orr-Sommerfeld equation with the Dedalus code \textcolor{black}{\cite{DedalusLib,DedalusWeb}} (where $\kxeq$ is the only length scale in the system) with a Reynolds number $Re = 400$. Their agreement with the other curves supports the notion that kinetic effects do not play a significant role in determining the growth rate of this mode. Crosses and dots in Fig.~\ref{dispersion} corresponding to lower values of $k_y$ show especially good agreement with the fluid results. As each curve represents a fixed $k_y$ with varying $\kxeq$, finite Larmor radius (FLR) effects become more important as $\kxeq$ increases (i.e.~as $k_y/\kxeq$ decreases), suggesting that the reduced growth rates in the $k_y = 0.2$ simulations relative to the fluid results are due to FLR effects. In non-periodic shear layers, such as $V \sim \tanh (x)$, it is observed that FLR effects can be stabilizing or destabilizing depending on the alignment of the equilibrium vorticity and magnetic field \cite{Faganello,Henri}. Due to the sinusoidal nature of the flow studied here, the simulation domain includes regions where vorticity and magnetic field are aligned and where they are anti-aligned, suggesting that the FLR stabilization observed in our system is qualitatively different from what is found in shear layers. We speculate that the FLR stabilization is due to a reduction in the gyro-averaged potential $\bar{\phi}$, as $\bar{\phi}/\phi$ generally decreases with increasing $k$.

\subsection{Forcing and dissipation terms}
In this paper, \textcolor{black}{nonlinear} calculations often include additional terms corresponding to forcing and dissipation, which we introduce here. Hyperdissipation $-\Dhyp(k_x^4 + k_y^4)\tilde{f}$ \textcolor{black}{\cite{MJCPC}} \textcolor{black}{is employed to provide} small-scale dissipation in place of collisions, which are not expected to sufficiently dissipate small-scale fluctuations at achievable resolutions within valid limits of collision models. We note that our hyperdissipation term differs from what is more standard in the fluids community, where $k_x^4 + k_y^4$ is replaced by $(k_x^2 + k_y^2)^2$. \textcolor{black}{A second dissipative term $-\Drad\tilde{f}$ is spatially uniform and sometimes referred to as radiative damping} \textcolor{black}{or friction \cite{TobiasFriction}}. It absorbs energy transferred to large scales \cite{TobiasFriction,Reynolds-Barredo}, while also serving as a ``symmetry-breaking'' parameter that adjusts the relative growth rates of linear modes without modifying their structure. 

Finally, we introduce a Krook operator $-\Dk \delta_{k_x, \pm \kxeq} \delta_{k_y, 0} \tilde{f}$, where $\delta_{i,j}$ is the Kronecker delta, to represent forcing of the unstable equilibrium and prevent it from decaying due to turbulent fluctuations \cite{PueschelKrook}. Aside from being linear in $f$ and therefore not constant in time, this is identical to the inhomogeneous body forcing used in studies of Kolmogorov flow \cite{Platt,Musacchio,Lucas}. While the sign of the Krook operator seems to suggest that it removes energy from the system, that is \textcolor{black}{merely} a consequence of our separation between equilibrium and perturbation. As explained in Ref.~\cite{Waleffe}, the kinetic energy of the full flow is $E = \int |\mathbf{V} + \mathbf{v}|^2 dx dy$, so that if the $(k_x, k_y) = (\kxeq, 0)$ component of $\mathbf{v}$ opposes that of $\mathbf{V}$ and is not larger in magnitude, as we will see to be the case in Sec.~IV, term\textcolor{black}{s} that appear to dissipate the ``perturbation energy" $\int |\mathbf{v}|^2 dx dy$ \textcolor{black}{at that wavenumber} will actually increase the true energy $E$.

\textcolor{black}{Having constructed a linear operator that yields consistent results for the most unstable eigenmode's growth rate, we now address the rest of the spectrum of eigenvalues.}


\section{Eigenspectrum}

\begin{figure}
	\includegraphics[width=16cm]{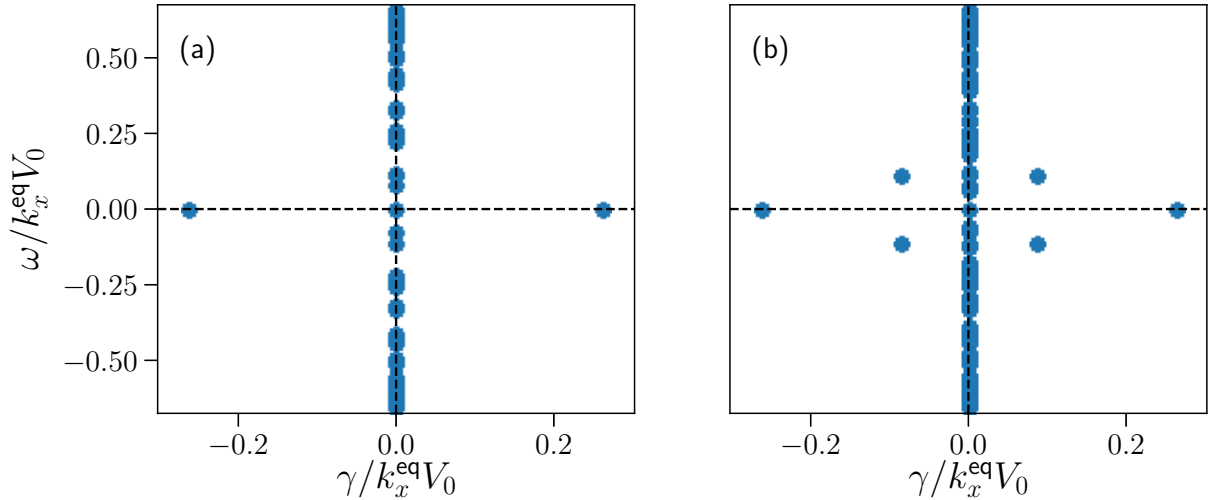}
	\caption{Eigenvalue spectra for $k_y/\kxeq = 2/3$ with $L_x = \lambda^\text{eq}$ (a) and $L_x = 2\lambda^\text{eq}$ (b). With $L_x = \lambda^\text{eq}$, at each unstable $k_y$, the spectrum includes one unstable and one stable mode with equal and opposite growth rate $\gamma$, as well as \textcolor{black}{a} continuous spectrum of marginal modes corresponding to resonances between the phase velocity $\omega / k_y$ and equilibrium flow \cite{Case}. As the box size is increased to fit multiple wavelengths of the equilibrium, additional stable and unstable modes are introduced \cite{Goodman,PessahGoodman}, and additional marginal eigenvalues appear due to an increase in number of values of $V_0 \cos (\kxeq x)$ sampled by the extended grid (thus additional resonances with $\omega / k_y$).}
	\label{spectra}
\end{figure}

\subsection{Subdominant modes}
At each $k_y$ there exists a spectrum of eigenmodes $\hat{f}_j$ and eigenvalues $\omega_j$, with corresponding potential structures $\hat{\phi}_j$. For $0 < k_y/\kxeq < 1$, we let $j=1$ denote the fastest-growing mode. The KH instability has been investigated in gyrokinetics before, but previous calculations did not address linear modes other than $\hat{f}_1$ \textcolor{black}{or their role in saturation}. With the linear operator $\mathcal{L}_\text{KH}$ now implemented in \textsc{Gene}, its full spectrum of eigenmodes and eigenvalues can be obtained \cite{GENE-EV,HatchLeft,TerryLeft}. \textcolor{black}{Like} the inviscid fluid analog, for each $k_y$ in the unstable range there exist one unstable mode, one stable (damped) mode with equal and opposite growth rate \cite{DrazinReid}, and a continuum of marginally stable modes \cite{Case}, shown in Fig.~\ref{spectra} (a) for $k_y/\kxeq = 2/3$. 
The additional degrees of freedom gained in gyrokinetics relative to a fluid calculation, by taking into account \textcolor{black}{the} velocity-space \textcolor{black}{structure} of multiple species, increases the rank of the discretized linear operator considerably, and leads to many more marginally stable modes. A single point on the marginally stable continuum in Fig.~\ref{spectra} corresponds to hundreds of eigenmodes (depending on velocity-space resolution), each with similar electrostatic potentials but significantly different velocity-space structure.

Despite the added degrees of freedom in gyrokinetics, there \textcolor{black}{are} still only \textcolor{black}{one} stable and \textcolor{black}{one} unstable eigenmode per $k_y$ for $0 < k_y/\kxeq < 1$ when the box size $L_x$ equals the wavelength of the equilibrium flow\textcolor{black}{, denoted by $\lambda^\text{eq} \equiv 2 \pi / \kxeq$}. Consistent with magnetohydrodynamic (MHD) studies of a similar system \cite{Goodman,PessahGoodman}, we find that \textcolor{black}{flows where} multiple wavelengths of the equilibrium \textcolor{black}{are present} (i.e.~setting $L_x = n\lambda^\text{eq}$ where $n \geq 2$ is an integer) \textcolor{black}{exhibit} pairs of subdominant unstable ($0 < \gamma_j < \gamma_1$) and stable \textcolor{black}{($\gamma_2 < \gamma_j < 0$)} modes, shown in Fig.~\ref{spectra} (b). This means that simulations with larger boxes but with an equilibrium flow of the same wavelength are expected to have different dynamics than simulations with $L_x = \lambda^\text{eq}$, as they include additional modes through which $\mathcal{L}_\text{KH}$ can inject or remove energy. \textcolor{black}{In Sec.~IV we will demonstrate that including $\Dk$ and $\Drad$ admits a system where, for sufficiently large $L_x$, observables are converged with respect to a further increase in $L_x$.}

\begin{figure}
	\includegraphics[width=16cm]{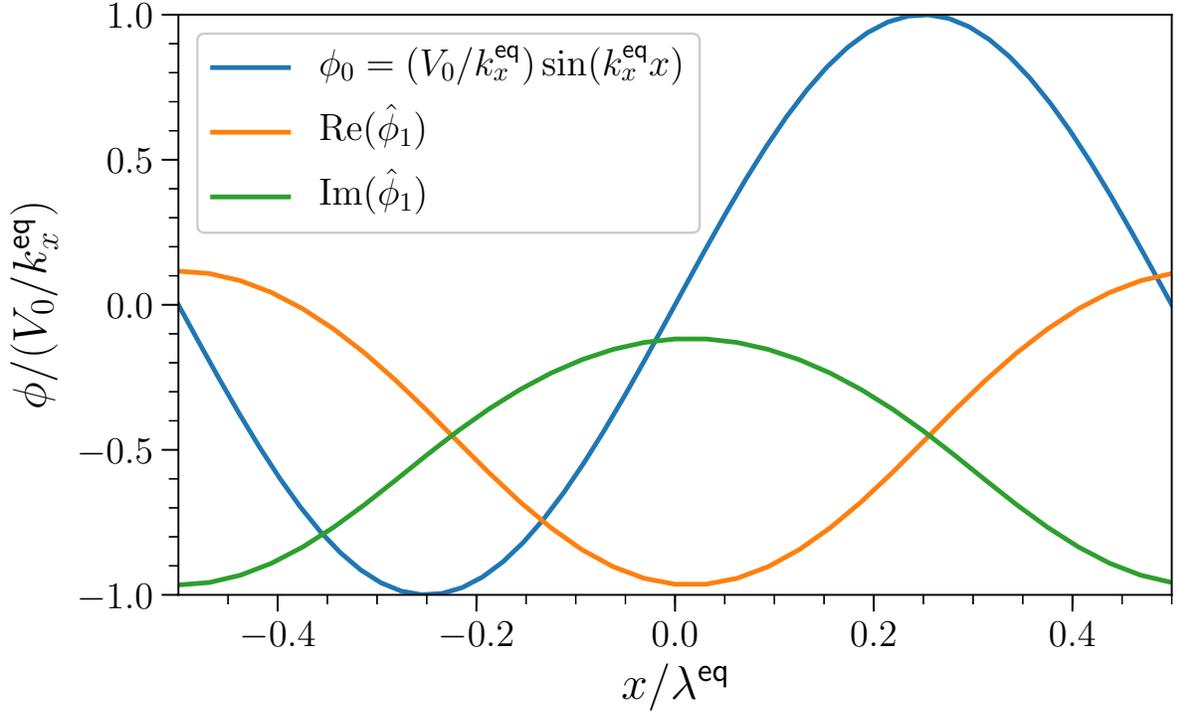}
	\caption{The equilibrium potential $\phi_0$ considered throughout this paper, which generates an $E \times B$ flow along the $y$-axis that varies sinusoidally in the $x$ direction with wavenumber $\kxeq$, alongside the real and imaginary parts of the potential corresponding to the unstable eigenmode $\hat{\phi}_1(x, k_y)$ plotted with respect to $x$ at $k_y/\kxeq = 1/2$. The stable eigenmode's potential is the complex conjugate of the unstable eigenmode's potential, $\hat{\phi}_2 = \hat{\phi}_1^*$.}
	\label{eigenmodes}
\end{figure}

As stated above, for each $k_y$ in $0 < k_y/\kxeq < 1$, we let $j=1$ refer to the \textcolor{black}{dominant} unstable mode. We will further let $j=2$ refer to the \textcolor{black}{corresponding} stable mode, and $j>2$ to all other modes. Figure \ref{eigenmodes} shows the $x$-dependence of $\hat{\phi}_1$ at $k_y/\kxeq = 1/2$ alongside the streamfunction for the equilibrium flow. Consistent with the fluid case \cite{Paper1,DrazinReid}, we find $\gamma_2 = -\gamma_1$ (such that both $|\gamma_{1,2}|$ are reduced by FLR effects), and $\hat{\phi}_2(x,k_y) = \hat{\phi}_1^*(x,k_y)$. Accordingly, we refer to $f_2$ as a conjugate stable mode.

Consistent with the fluid case discussed in Sec.~II.~B, the sinusoidal nature of the equilibrium \textcolor{black}{gives} eigenmodes a \textcolor{black}{discrete,} comb-like structure in $k_x$, where $\tilde{f}_j$ is zero at every $k_x$ except for a countably infinite number of $k_x$ that are each separated by $\kxeq$. All of the modes whose eigenvalues are plotted in Fig.~\ref{spectra} (a), including $\tilde{f}_1$ and $\tilde{f}_2$, have nonzero amplitudes at integer multiples of $\kxeq$. Many of the additional modes gained in Fig.~\ref{spectra} (b) by extending $L_x$ to $2\lambda^\text{eq}$, such as the modes with finite growth rate and real frequency, are nonzero at half-integer multiples of $\kxeq$. This implies that arbitrary linear combinations of the modes in Fig.~\ref{spectra} (a) are only nonzero at integer multiples of $\kxeq$.

\subsection{Forcing and dissipation effects}
The additional physics effects introduced in Sec.~II.~D each modify the eigenmodes to varying degrees. The Krook operator enters the Vlasov equation \textcolor{black}{only} at $k_y = 0$, so it has no impact on the $k_y > 0$ \textcolor{black}{eigenmode spectra}. The radiative damping term reduces the growth rate of every eigenmode by $D_\text{rad}$ without changing the mode structure. \textcolor{black}{T}he hyperdissipation term has a more significant impact on the spectrum. It reduces the growth rate of the unstable mode with minor modifications to its structure, and replaces both the stable mode and marginal continuum with a set of damped modes that does not include any mode \textcolor{black}{resembling} the conjugate stable mode.

\textcolor{black}{We now turn our attention the nonlinear saturation of this system.}

\section{Instability saturation}

\begin{figure}
\includegraphics[width=16cm]{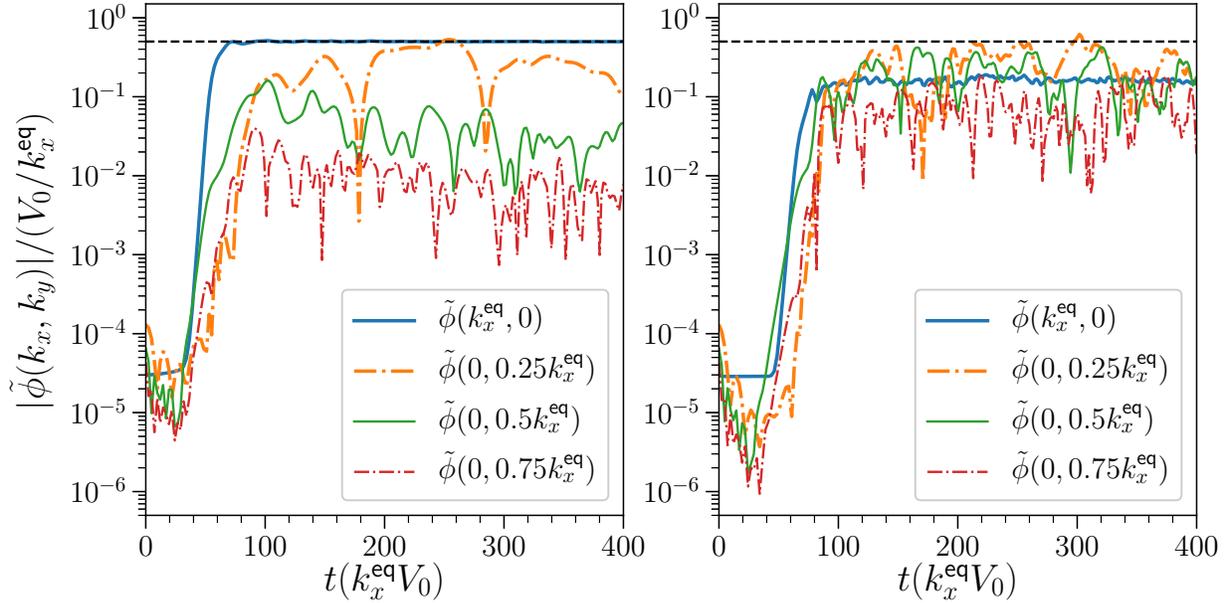}
\caption{Left: Nonlinear simulations with dissipation quasilinearly flatten\textcolor{black}{,} then decay. Quasilinear flattening is measured by investigating $\tilde{\phi}$ at $(k_x,k_y) = (\pm \kxeq, 0)$. The perturbation cancels the drive once $\tilde{\phi}(\kxeq,0)$ (blue) reaches a magnitude of $0.5 V_0/\kxeq$ \textcolor{black}{(black dashed line)}. Linearly unstable Fourier modes then turbulently decay over time. Right: Introducing a Krook operator ($\Dk/(\kxeq V_0) = 1$ here) \textcolor{black}{partially} suppresses the Fourier mode responsible for quasilinear flattening, driving the system and leading to a quasi-stationary state of driven turbulence.}
\label{phi timetraces}
\end{figure}

\subsection{Saturation and decaying turbulence}
To investigate the saturation of this instability, we \textcolor{black}{include in} Eq.~\eqref{Linear GK eqn} the full $E \times B$ nonlinearity, yielding
\begin{equation}\label{NonlinearEq}
\frac{\partial f}{\partial t} = \mathcal{L}_\text{KH}[f] + \left\{ f, \bar{\phi}\right\}.
\end{equation}
Owing to the way in which the linear drive terms were derived and implemented, the evolution of Eq.~\eqref{NonlinearEq} with some initial condition $f_\text{init}$ is identical to the evolution of Eq.~\eqref{RogersEq} with the initial condition $f_0 + f_\text{init}$\textcolor{black}{, presuming no dissipation or drive is included}. When dissipation terms are added to Eq.~\eqref{NonlinearEq}\textcolor{black}{, they} do not act on $f_0$\textcolor{black}{, unlike those in Eq.~\eqref{RogersEq}}.

\begin{figure}
\includegraphics[width=16cm]{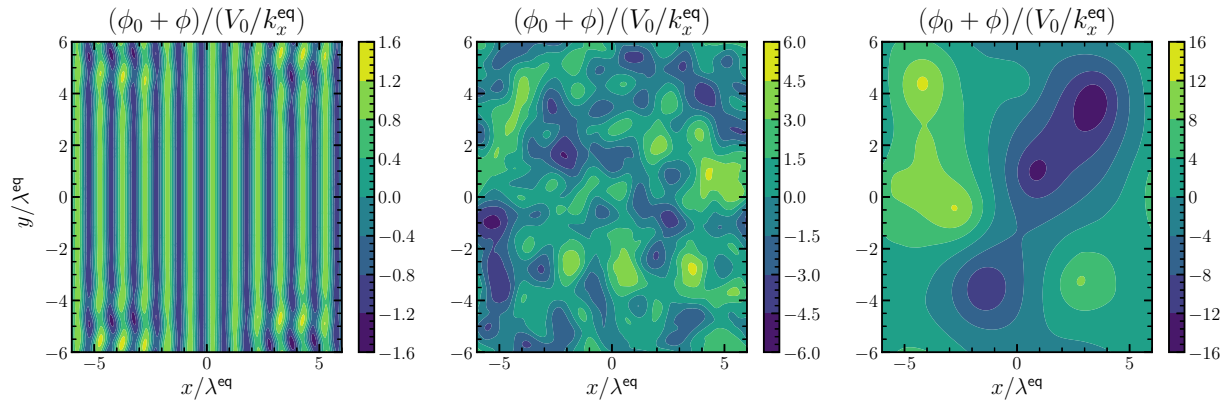}
\caption{Contours of the full (equilibrium and fluctuation) electrostatic potential for a nonlinear simulation with $D_\text{Krook} = 0$. From left to right, plots correspond to $t (\kxeq V_0) \approx 46, 101$, and $502$. Center and right plots show the tendency for small-scale fluctuations to dissipate, leaving coherent vortices that merge to progressively larger scales.}
\label{decaying contours}
\end{figure}

\textcolor{black}{As the system evolves according to Eq.~\eqref{NonlinearEq},} the nonlinearity transfers energy across a range of scales, but with zero energy injection and nonzero dissipation, the initial energy eventually decays away. In terms of saturation of a linear instability, this can be understood as quasilinear flattening, where the fluctuations \textcolor{black}{reduce mean} gradients until the system is linearly stable. This is observed in simulations of Eq.~\eqref{NonlinearEq} with added hyperdissipation, as shown in Figs.~\ref{phi timetraces} and \ref{decaying contours}. Once unstable wavenumbers reach a sufficient amplitude, fluctuation\textcolor{black}{s} at the wavenumber\textcolor{black}{s} of the equilibrium flow, \textcolor{black}{i.e.} $(k_x, k_y) = (\pm \kxeq, 0)$, quickly grow to \textcolor{black}{offset the unstable profile of the mean flow}. \textcolor{black}{From} this point the system exhibits features of decaying turbulence: the dynamics are highly intermittent, with long periods of coherent behavior punctuated by the merging of vortices\textcolor{black}{.} This is consistent with previous 2D KH simulations \cite{Faganello}, and can be expected given the lack of external forcing; the linear drive in Eq.~\eqref{NonlinearEq} appears similar to an external forcing term, but as argued in the preceding paragraph, that is merely a consequence of the formal separation between the equilibrium and fluctuations.

\subsection{Driven turbulence}

In \textcolor{black}{many} physical systems where shear-flow instability saturation and turbulence are of interest, the unstable shear flow is not \textcolor{black}{some} ideal initial condition but is brought about by a separate process. \textcolor{black}{E}xamples include shear flows \textcolor{black}{driven} by boundary conditions \cite{Brandstater}\textcolor{black}{,} drift-wave instabilities \cite{Rogers2000,Rogers2005} in laboratory experiments, and jets, gravity, or another instability \cite{Goodman} in astrophysical systems.
We include a Krook operator, introduced in Sec.~II, with the intent of capturing some of the effects of such \textcolor{black}{continual} forcing but without modeling the subtleties of any particular system where forcing produces a shear flow.

\begin{figure}
	\includegraphics[width=16cm]{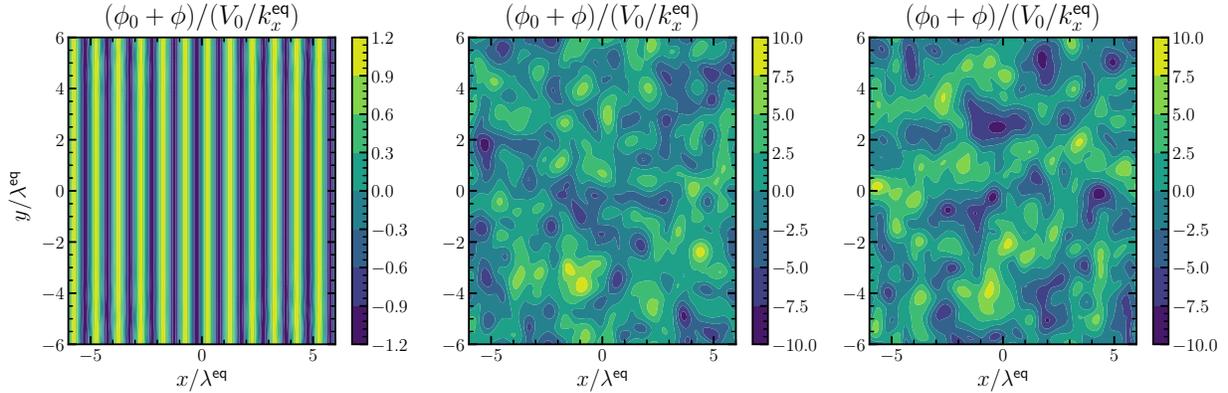}
	\caption{Contours of the full (equilibrium and fluctuation) electrostatic potential for a nonlinear simulation with $\Dk/(\kxeq V_0) = 1$ and $\Drad/(\kxeq V_0) = 0.05$. From left to right, plots correspond to $t (\kxeq V_0) \approx 46, 102$, and $501$. Comparing with Fig.~\ref{decaying contours} shows the system no longer tends towards large-scale coherent vortices with gradual decay of energy. Instead, multiple scales are excited \textcolor{black}{and form a} quasi-stationary state.}
	\label{driven contours}
\end{figure}

The result of including the Krook operator is \textcolor{black}{readily} seen in Figs.~\ref{phi timetraces} and \ref{driven contours}. When the Krook operator is added to Eq.~\eqref{NonlinearEq}, it suppresses the tendency for the $(k_x, k_y) = (\pm \kxeq, 0)$ component of the fluctuation to cancel out $f_0$, \textcolor{black}{thereby} injecting energy into the system by reinforcing the unstable equilibrium. This in turn drives other Fourier modes via the KH instability\textcolor{black}{, as} is seen in the timetrace of $\tilde{\phi}(\kxeq, 0)$, shown in Fig.~\ref{phi timetraces}: the $(k_x, k_y) = (\pm \kxeq, 0)$ component no longer reaches the amplitude necessary to cancel out the driving shear flow, and other Fourier modes no longer decay over time, leading to a quasi-stationary state of driven turbulence where the energy injected by the Krook drive is balanced by energy dissipation. As $\Dk$ increases, the saturated amplitude of $\tilde{\phi}(\pm \kxeq, 0)$ decreases, corresponding to an overall increase in $\tilde{\phi}_0 + \tilde{\phi}(\pm \kxeq, 0)$. The dominant balance that determines the amplitude of $\tilde{\phi}(\pm \kxeq, 0)$ in saturation is between the Krook drive and the Reynolds stress, which we explore further in Sec.~IV.~C.

Also observed in Fig.~\ref{decaying contours} is the tendency for the system to form coherent vortices that gradually merge to the largest scale allowed by the simulation domain. \textcolor{black}{This behavior is also} observed in 2D shear layer simulations \cite{Faganello}, \textcolor{black}{and is} consistent with the inverse \textcolor{black}{energy} cascade to large scales in 2D hydrodynamics. \textcolor{black}{The inverse cascade leads} to a system \textcolor{black}{with} saturation properties \textcolor{black}{that} change as the box size is increased. The radiative damping term $\Drad$ introduced in Sec.~II damps low-$k$ fluctuations, preventing energy from continuously building up at the largest scales, and thereby allowing fluctuation spectra to reach a stationary condition at low $k$. For this reason\textcolor{black}{, and for the sake of presenting simulations where observables are converged with respect to the box size}, the majority of our simulations were run with $L_x = 12 \lambda^\text{eq}$ and $\Drad = 0.05$, \textcolor{black}{a rate that is} roughly $20\%$ of the maximum linear growth rate in the dissipationless case. Figure \ref{driven contours} shows the results of a simulation with these parameters \textcolor{black}{and} $\Dk = 1$\textcolor{black}{. In contrast with} Fig.~\ref{decaying contours}\textcolor{black}{, t}he system exhibits multiple excited scales in a quasi-stationary saturated state\textcolor{black}{, providing the type of turbulence desired for studying momentum transport and eigenmode excitation.}

\subsection{Momentum transport}
We \textcolor{black}{investigate} the momentum transport driven by turbulent fluctuations in this system\textcolor{black}{, examining} the $xy$ component of the Reynolds stress tensor, denoted as $\tau$\textcolor{black}{. From the average of} the product of the $x$ and $y$ components of the fluctuating $E \times B$ flow in the homogeneous $y$ direction,
\begin{equation}\label{tau def}
\tau \equiv \left\langle - \frac{\partial\phi}{\partial x} \frac{\partial\phi}{\partial y} \right\rangle_y,
\end{equation} 
where $\langle A \rangle_q$ denotes an average of some quantity $A$ over a domain in the variable $q$. Due to the sinusoidal variation in $x$ of the equilibrium, $\tau$ changes sign along the $x$ axis as the sign of the equilibrium flow changes, an expected feature of Kolmogorov flow \cite{Musacchio}.

In \textcolor{black}{nonlinear} gyrokinetic \textcolor{black}{simulations}, numerical convergence is typically tested by measuring changes of some scalar, time-averaged transport quantity with \textcolor{black}{numerical parameters such as} resolution and domain size.
Due to the changes in sign of $\tau$, the average of $\tau$ in the $x$ direction and time, $\langle \tau \rangle_{x,t}$, is not appropriate for testing numerical convergence because it is typically  $0$.
Instead, we calculate the root-mean-square (RMS) of $\tau$, i.e., $\sqrt{\langle \tau^2 \rangle_x}$, and compare the time-average in the quasi-stationary state as resolution changes.
For the simulation shown in Fig.~\ref{driven contours}, the time-averaged $\tau_\text{RMS}$ in saturation changes by at most $2\%$ when any \textcolor{black}{spatial or velocity} coordinate's domain size or resolution is doubled except $L_x$ (expected due to the subdominant unstable modes and inverse cascade), where it changes by $9\%$. Therefore, despite creating additional unstable and stable eigenmodes as box size is increased, this simulation is numerically converged in $L_x$ with regards to $\tau_\text{RMS}$.

\begin{figure}
	\includegraphics[width=8cm]{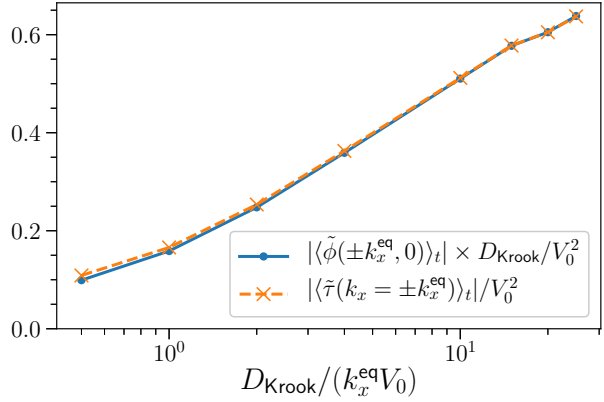}
	\caption{Comparison between average Krook drive amplitude $|\langle \tilde{\phi}(\pm k_x^\text{eq},0) \rangle_t | \Dk$ and the average amplitude of the corresponding Fourier component of the Reynolds stress $\tau$ in saturation across a range of driving frequencies $\Dk$. Other simulation parameters are the same as in Fig.~\ref{driven contours}. In the saturated state, the mean flow is governed by a competition between the external forcing and the turbulent Reynolds stress. A small contribution is made by the influence of dissipation on $\tilde{\phi}(\pm k_x^\text{eq},0)$, evidenced by the minor mismatch between the two curves at the lowest values of $\Dk$.}
	\label{force balance}
\end{figure}

Consistent with studies of Kolmogorov flow (where a constant force is typically used, while our forcing is proportional to $\tilde{f}(\kxeq, 0)$) \cite{Musacchio}, we find that as the forcing increases, both the mean flow velocity and the Reynolds stress increase, such that at saturation the two are in balance. This is shown in Fig.~\ref{force balance}, where the force on the mean flow applied by $\Dk$ is seen to balance the force due to $\tau$. This can also be seen by considering the effect of a similar Krook operator on Eq.~\eqref{nonlinear vorticity}. When Eq.~\eqref{nonlinear vorticity} is Fourier transformed in both $x$ and $y$, our forcing term appears as $\Dk \delta_{k_x, \pm \kxeq} \delta_{k_y, 0} (k_x^2 + k_y^2) \tilde{\phi}$. The $(k_x, k_y) = (\kxeq, 0)$ component of the equation then becomes
\begin{equation}\label{mean nonlinear vorticity}
\frac{\partial}{\partial t}\tilde{\phi}(\kxeq,0) + \sum_{\mathbf{k}'} \frac{k'_y}{\kxeq}\left[(\kxeq - k'_x)^2 + k'^2_y\right]\tilde{\phi}(k'_x, k'_y)\tilde{\phi}(\kxeq - k'_x, -k'_y) = - \Dk \tilde{\phi}(\kxeq,0),
\end{equation}
where the nonlinear term is the $k_x = \kxeq$ component of the Fourier-transformed Reynolds stress $\tilde{\tau}$. For a quasi-stationary, saturated state, the time-average of Eq.~\eqref{mean nonlinear vorticity} yields a balance between the Reynolds stress and the $\Dk$ term. Figure \ref{force balance} compares these terms for a range of simulations with different values of $\Dk$, demonstrating good agreement with expectations. A small mismatch occurs because the effect of dissipation on the flow makes a small contribution to the force balance, but the other forces are clearly dominant. Because we only include dissipation on the fluctuation, not the equilibrium $\phi_0$ which is independent of $\Dk$, this contribution decreases as $\Dk$ increases.

\section{Eigenmode analysis}

\subsection{Eigenmode Expansion}
We investigate the role of stable modes in this system by expanding the turbulent state in a basis of the eigenmodes of the dissipationless operator $\mathcal{L}_\text{KH}$. We expand in eigenmodes of the dissipationless operator to focus on the role played by $f_2$, which vanishes in the dissipative system. This also allows for comparison with previous work \cite{Paper1}, where the dissipationless modes were considered.

\begin{figure}
	\includegraphics[width=16cm]{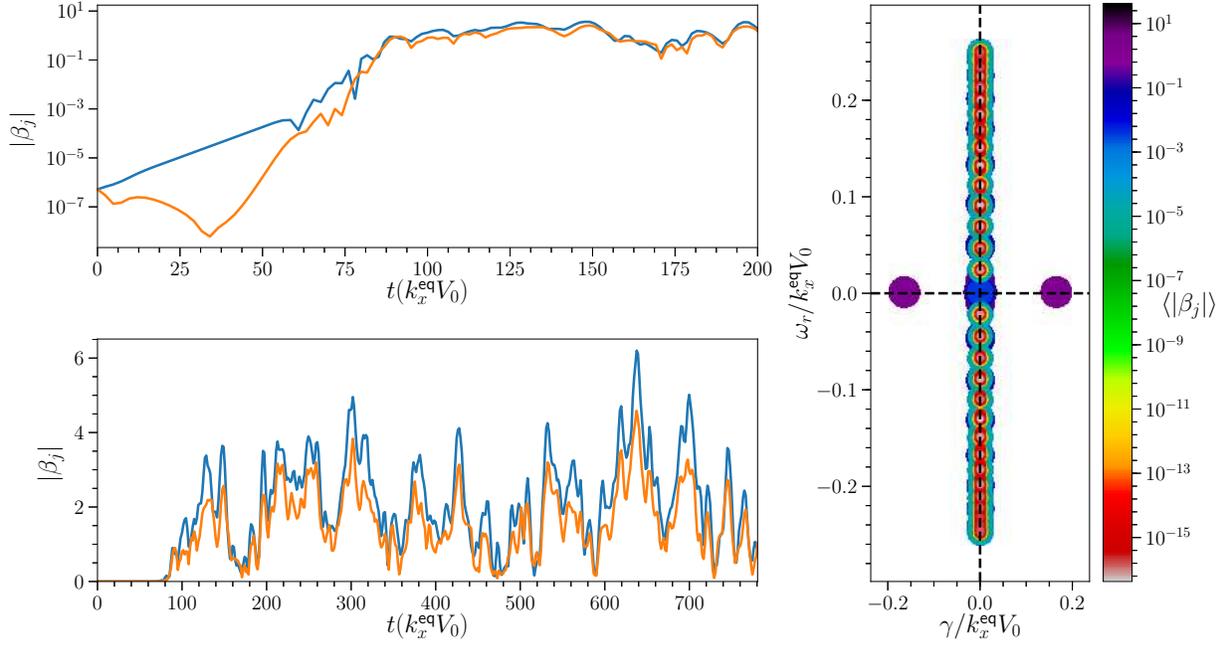}
	\caption{Left: Amplitudes of the unstable and stable eigenmodes, $|\beta_1|$ and $|\beta_2|$\textcolor{black}{, respectively}, as function\textcolor{black}{s} of time on a logarithmic scale (top, horizontal axis reduced to highlight the parametric growth of $|\beta_2|$) and a linear scale (bottom) at horizontal wavenumber $k_y/\kxeq = 0.25$ for the simulation with $\Dk = 1$, $\Drad = 0.05$, and $\Dhyp = 1.6$. Right: Full spectrum of eigenvalues $\omega_j$ (real part \textcolor{black}{$\omega_r$} on $y$-axis, growth rate \textcolor{black}{$\gamma$} on $x$-axis), with color indicating time-averaged (starting from $t = 300$) amplitudes $\langle |\beta_j| \rangle$, and dot size scaled proportionally to allow multiple $|\beta_j|$ with the same $\omega_j$ to be shown. The decay of $\beta_2$ \textcolor{black}{is} followed by nonlinear growth much faster than $\beta_1$\textcolor{black}{,} while $\beta_1$ continues its linear growth\textcolor{black}{,} consistent with Ref.~\cite{Paper1}. The remarkabl\textcolor{black}{e} similar\textcolor{black}{ity of the} values of $|\beta_1|$ and $|\beta_2|$ in the saturated state was \textcolor{black}{predicted in} Ref.~\cite{Paper1}. Results are qualitatively similar for other unstable $k_y$.}
	\label{base case betas}
\end{figure}

As discussed in Sec.~III, the operator $\mathcal{L}_\text{KH}$ has a distinct set of $N_\mathrm{ev}$ eigenmodes $\{ \hat{f}_j \}$ for each value of $k_y$. Therefore, an expansion of an arbitrary state $f(s, x, y, v_\parallel, \mu)$ in a basis of eigenmodes $\hat{f}_j$ may be written as
\begin{equation}\label{eigenmode expansion}
f(s, x, y, v_\parallel, \mu) = \sum_{k_y} \sum_{j=1}^{N_\mathrm{ev}} \beta_j(k_y) \hat{f}_j(s, x, k_y, v_\parallel, \mu) e^{i k_y y}.
\end{equation}
As in Sec.~III, the index $j$ is a positive integer that enumerates the eigenmodes at a given $k_y$, and for $0 < k_y / \kxeq < 1, j=1$ \textcolor{black}{and $j=2$ label} the most unstable mode and its stable conjugate\textcolor{black}{, respectively}. The number of eigenmodes $N_{ev}$ obtained by the eigenmode solver is equal to the number of degrees of freedom in the discrete numerical representation, i.e., the product of the number of grid points and the number of species, and the modes were verified to be linearly independent, so expansions of this form exist and are unique assuming the numerical resolutions of both sides of Eq.~\eqref{eigenmode expansion} are identical. Figure \ref{base case betas} shows timetraces of $|\beta_1|$ and $|\beta_2|$, as well as the time-averaged $|\beta_j|$ in saturation for every $j$ at $k_y/\kxeq = 0.25$ for the same simulation shown in Fig.~\ref{driven contours}.

The values $\beta_j$, which we refer to as the amplitudes of each eigenmode, can be understood as coordinates or components of $f$ in the basis $\{ f_j \}$. When such an expansion is performed at multiple time steps of a given simulation so that $f$ is a function of time, each $\beta_j$ becomes a function of time that indicates the relative contribution of eigenmode $f_j$ to the state of the system over time. In linear simulations, $\beta_j(k_y, t) = \beta_j(k_y, 0) e^{i \omega_j t}$ for each $j$ and $k_y$. Previous work showed how $\beta_1$ and $\beta_2$ interact nonlinearly in a \textcolor{black}{fluid} system, derived equations for $\partial\beta_j/\partial t$ by inserting expansions of the form Eq.~\eqref{eigenmode expansion} into the governing equations of the system, and compared the relative sizes of different terms leading into instability saturation \cite{Terry2006,Makwana,Paper1}. Here we directly calculate the evolution of each $\beta_j$ over time in nonlinear simulations\textcolor{black}{, extending analysis beyond the onset of saturation}. Our procedure for calculating each $\beta_j$ relies on the left eigenmodes of $\mathcal{L}_\text{KH}$ and is described in Refs.~\textcolor{black}{\cite{TerryLeft,HatchLeft}}.

Similar analyses have been performed for \textcolor{black}{gyroradius-scale instabilities in} reduced fluid models \cite{Makwana}, and gyrokinetic \textcolor{black}{models} \cite{HatchLeft,MJBen}. These eigenmode expansions are \textcolor{black}{related} to the ``projections" calculated in related work \cite{Whelan,MJBen,Ben2018}, defined as 
\begin{equation}\label{Projection}
p_j = \left|\int d\mathbf{x} d\mathbf{v} \sum_s f_j^* f_\mathrm{NL}\right| \left(\int d\mathbf{x} d\mathbf{v} \sum_s |f_j|^2 \int d\mathbf{x} d\mathbf{v} \sum_s |f_\mathrm{NL}|^2\right)^{-1/2}
\end{equation}
(where $f_\mathrm{NL}$ is the \textcolor{black}{nonlinearly-evolved} distribution function, and \textcolor{black}{the summations are over each species}), but the two are generally quite different. Identifying $\langle g, h \rangle \equiv \int d\mathbf{x} d\mathbf{v} g^* h$ as an inner product on the space of distribution functions $f$, projections $p_j$ are inner products normalized by the lengths of $f_j$ and $f$ so that $p_j=0$ if they are orthogonal (under this inner product) and $p_j=1$ if they are parallel. The eigenvectors of an arbitrary linear operator are not guaranteed to be mutually orthogonal under a given inner product (we have verified that the eigenmodes of our system are not mutually orthogonal under the above inner product), which leads to the possibility \textcolor{black}{that} the projection onto one eigenvector depend\textcolor{black}{s} on the amplitudes of every eigenvector. For example, one could find that the projection $p_j$ onto a stable mode counterintuitively grows over time in a linear simulation due to nonorthogonality, even though the amplitude $\beta_j$ of the stable mode decays. Likewise, if the projection onto a stable mode is large in \textcolor{black}{the} saturated state, it is not immediately clear whether this is due to a large stable mode amplitude, significant nonorthogonality with the dominant unstable mode, or even due to nonorthogonality with an entirely different mode that has a large amplitude. This situation is avoided if the linear operator has mutually orthogonal eigenvectors (e.g.~if it is Hermitian), if the set of modes $f_j$ are replaced by an orthogonal set, such as from a proper orthogonal decomposition \cite{HatchLeft}\textcolor{black}{,} or by applying an orthogonalization procedure \textcolor{black}{like} Gram-Schmidt \cite{MJBen}. However, the relationship between the eigenmode amplitudes and the orthogonalized mode amplitudes is not immediately clear. We focus our attention on the eigenmode amplitudes $\beta_j$ rather than projections $p$ because linear energy transfer due to $\mathcal{L}_\text{KH}$ is directly related to $\beta_j$, not $p$ \cite{Terry2006}, and to facilitate comparison with Ref.~\cite{Paper1}.

For the simulation shown in Fig.~\ref{driven contours}, we use the parameters $L_x = 12\lambda^\text{eq}, \Dk = 1, \Drad = 0.05$, and $\Dhyp = 1.6$, with $512$ grid points in the $x$ direction. Calculating every eigenmode \textcolor{black}{of the system at that resolution is prohibitively expensive}. Instead, to generate Fig.~\ref{base case betas} we perform eigenvalue computations with $L_x = \lambda^\text{eq}$. Due to the \textcolor{black}{discrete,} comb-like \textcolor{black}{eigenmode} structure in $k_x$ discussed in Secs.~II.~B and III.~B, this reduced set of modes does not describe the full state \textcolor{black}{of} Eq.~\eqref{eigenmode expansion} because \textcolor{black}{it lacks} modes obtained when $L_x > \lambda^\text{eq}$ [see Fig.~\ref{spectra}]. But this does allow for a full expansion of the components of $\tilde{f}(k_x, k_y)$ given by $k_x = n \kxeq$ for integer $n$, and this does not affect the obtained values of $\beta_1$ and $\beta_2$.

Consistent with analytical calculations and reduced models \cite{Terry2006,Makwana,Paper1}, Fig.~\ref{base case betas} shows that $|\beta_2|$ decays before being nonlinearly driven at a rate faster than the unstable mode's concurrent exponential growth. We stress that the evolution of $|\beta_2|$ is remarkably consistent with the inviscid \textcolor{black}{fluid problem} \cite{Paper1} despite the influence of nonzero $\Dhyp$ in the nonlinear simulation, which modifies the structure of $f_1$ and eliminates the conjugate stable mode $f_2$ from the eigenmode spectrum of the dissipative operator. A similar observation was made in studies of ITG pseudospectra, where a similar conjugate stable mode vanished in the dissipative case, but was nonetheless a part of the pseudospectrum and \textcolor{black}{was significantly} excited in saturation \cite{HatchPseudo}. Figure \ref{base case betas} only shows amplitudes for the $k_y/\kxeq = 0.25$ eigenmodes, but every other $k_y$ in $0 < k_y / \kxeq < 1$ exhibits similar results. The amplitude of $f_2$ in saturation nearly matches that of $f_1$ both at saturation onset and for the rest of the simulation. Since the two modes are nearly conjugate symmetric, this \textcolor{black}{suggests} that the linear energy dissipation due to $f_2$ \textcolor{black}{is a significant fraction of} the linear energy injection due to $f_1$ at the onset of saturation and throughout the quasi-stationary state\textcolor{black}{. This suggests} that the predictive capabilities of the \textcolor{black}{threshold parameter} $P_t$ analysis \textcolor{black}{studied in Refs.~\cite{Terry2006,Makwana}} carry over to systems \textcolor{black}{more general} than plasma microturbulence, and that a significant amount of the energy transferred to $k_y>0$ fluctuations via $\mathcal{L}_\text{KH}$ makes its way back into the mean flow rather than smaller scales.

\subsection{Truncated eigenmode expansions}
In turbulence models, it is common practice to separate the flow into mean and fluctuat\textcolor{black}{ing parts. If there is further separation between large and small scale structures, the former are often approximated by the most unstable eigenmode} \cite{Gaster,Liou,Nikitopoulos}. Here we demonstrate the potential for \textcolor{black}{improving} such models by including the stable mode in the approximation for the large scales.

\begin{figure}
	\includegraphics[width=8cm]{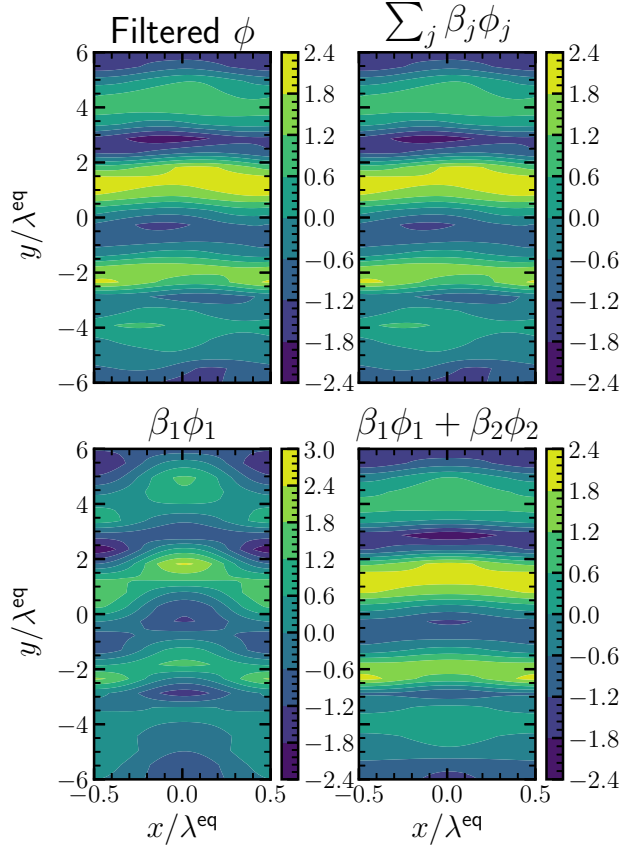}
	\caption{Comparison between the components of $\phi$ that are spanned by the eigenmodes in Fig.~\ref{base case betas} (top left), a summation over all of the eigenmodes in Fig.~\ref{base case betas} at every unstable $k_y$ (top right), summation over just the most unstable mode at every unstable $k_y$ (bottom left), and summation over the most unstable and conjugate stable mode at every unstable $k_y$ (bottom right) at $t \approx 501 (V_0 \kxeq)^{-1}$ for the same simulation as Fig.~\ref{driven contours}. Due to only integer multiples of $\kxeq$ contributing to these eigenmodes, they are unable to effectively reproduce the full flow profile, plotted in Fig.~\ref{driven contours}. However, those components of $\phi$ that can be expressed as a linear combination of the eigenmodes in Fig.~\ref{base case betas} are very well-described even by just the unstable $\phi_1$ and stable $\phi_2$.}
	\label{reduced phi contours}
\end{figure}

Figure \ref{reduced phi contours} compares part of the flow structure at $t \approx 501 (V_0 \kxeq)^{-1}$ to three different expansions. The top-left contours show the electrostatic potential $\phi$ for the simulation described in Fig.~\ref{base case betas}. To focus on the components of $\phi$ where the eigenmodes discussed in Figs.~\ref{dispersion} and \ref{spectra} can be used to approximate the flow, \textcolor{black}{a filtering procedure has been applied in Fig.~\ref{reduced phi contours} to remove} all but a subset of Fourier components $(k_x, k_y)$ have been artificially removed. Only $k_y$ in $0 < k_y / \kxeq < 1$ are included, and only $k_x = n \kxeq$ for integer $n$ are included. This allows the eigenmodes in Fig.~\ref{base case betas}, and the equivalent eigenmodes at other unstable $k_y$, to be used as a basis in the sense of Eq.~\eqref{eigenmode expansion}. The top-right contours show the $\phi$ structure obtained from summing over these eigenmodes at each unstable $k_y$, verifying that they indeed serve as a basis. The excellent agreement helps demonstrate that the wavenumber filtering only affects the amplitudes $\beta_j$ of eigenmodes that arise \textcolor{black}{from} having $L_x > \lambda^\text{eq}$, and fully captures the structure and amplitudes of the $L_x = \lambda^\text{eq}$ eigenmodes. Extremely minor differences between the top-left and top-right contours arise due to the higher $x$ resolution in the nonlinear simulation than in the linear eigenmode calculations. To investigate the differences between these large-scale flows and the results of approximating them using just the unstable mode, the bottom-left contours show the result of excluding every eigenmode in Eq.~\eqref{eigenmode expansion} except the most unstable at each $k_y$, as is often done in reduced models. The bottom-right contours are obtained similarly, but both the most unstable mode $\phi_1$ and the conjugate stable mode $\phi_2$ at each $k_y$ are included. Including $\phi_2$ produces a flow structure that is remarkably similar to the top-left and top-right flow structures, unlike what one obtains when only $\phi_1$ is included. Unsurprisingly, the more accurate flow structure leads to a more accurate Reynolds stress \textcolor{black}{(not shown)}.

\begin{figure}
	\includegraphics[width=8cm]{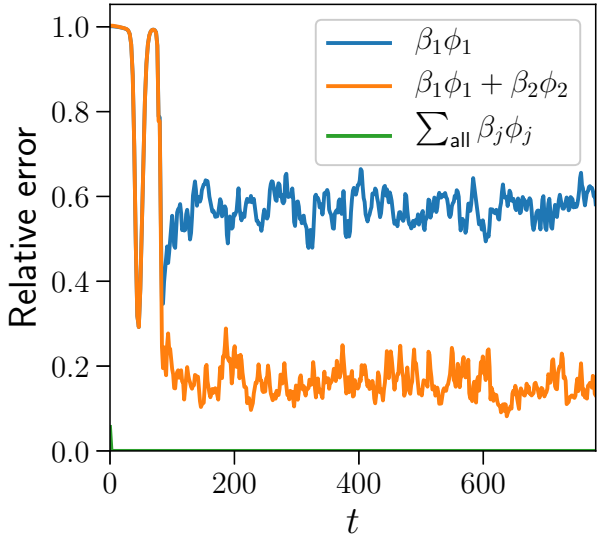}
	\caption{Error in $\phi$ of each of the eigenmode expansions of Fig.~\ref{reduced phi contours} relative to the filtered nonlinear data. \textcolor{black}{I}ncluding $\phi_2$ significantly improves fluctuation estimates \textcolor{black}{in the quasi-stationary state}.}
	\label{phi error}
\end{figure}

To compare the efficacy of these three eigenmode expansions over time, rather than the one timestep shown in Fig.~\ref{reduced phi contours}, we calculate the error \textcolor{black}{$\text{error} = ||\phi - \sum_j \beta_j \phi_j || / ||\phi||$} of each relative to the filtered nonlinear data (the top-left plot in Fig.~\ref{reduced phi contours})\textcolor{black}{. Here} $\phi$ refers to the filtered nonlinear $\phi$, and $||.||$ is the standard $L_2$ norm. Due to differences in $x$ resolution, the full expansion (in green) has minor errors that decay away as the simulation progresses. Errors in both the unstable-only expansion (blue) and the combined unstable-stable expansion (orange) start large due to choice of initial condition, gradually decay as the most unstable mode grows in the linear phase, and peak at the onset of saturation before fluctuating about an average value in the quasi-stationary state, with the inclusion of the stable mode reducing the average error in the saturated state by a factor of \textcolor{black}{three}.

\subsection{Influence of forcing and dissipation}

\begin{figure}
	\includegraphics[width=8cm]{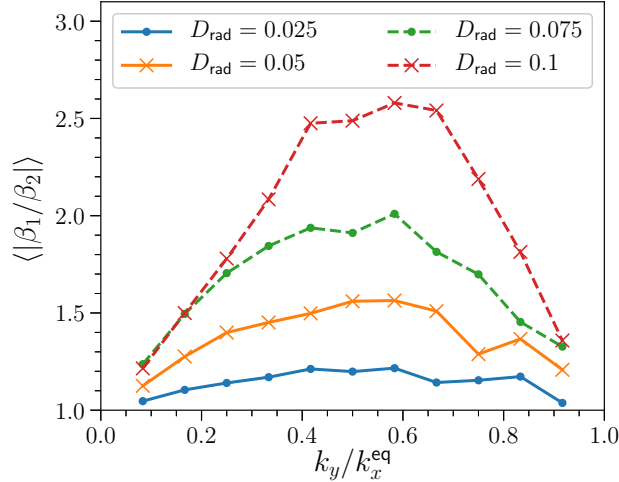}
	\caption{Time-averaged ratio of unstable mode amplitude to stable mode amplitude in saturation at each $k_y$ for a range of $\Drad$ \textcolor{black}{with $\Dk = 1$}. The growth rate dependence of the $P_t$ analysis \cite{Terry2006,Terry2009,Makwana} suggests that higher $\Drad$ causes $\beta_2$ to be driven less leading into saturation. Here we see that this is reflected in the eigenmode amplitudes in the saturated state. The $k_y$ dependence of the ratio $|\beta_1/\beta_2|$ roughly follows that of $\gamma_1$.}
	\label{rad scan}
\end{figure}

Figure \ref{base case betas} shows significant excitation of the stable mode in the saturated state for a simulation with $\Dk = 1$, $\Drad = 0.05$, and $\Dhyp = 1.6$, with Figs.~\ref{reduced phi contours} and \ref{phi error} demonstrating its importance in describing the large-scale fluctuations in $\phi$. To investigate the role of these parameters in determining the influence of stable modes in saturation, we vary them between different simulations. In particular, we pay close attention to the relative amplitudes of $\beta_1$ and $\beta_2$ as $\Drad$ and $\Dk$ are varied. Because $\Drad$ is a symmetry-breaking term in the sense that it decreases the growth rate of $f_1$ and increases the damping of $f_2$ without changing their mode structures, it reduces the parametric driving of $f_2$ by $f_1$\textcolor{black}{. (T}he parametric driving of $f_2$ by $f_1$ depends on their mode structures, the form of the nonlinearity, and $\gamma_2/\gamma_1$ \cite{Terry2006,Terry2009,Makwana,Paper1}\textcolor{black}{. O}f those, only $\gamma_2/\gamma_1$ is affected by $\Drad$, making its influence on $|\beta_1/\beta_2|$ more transparent\textcolor{black}{.)} Figure \ref{rad scan} shows that this leads to significantly smaller $|\beta_2|$ relative to $|\beta_1|$ in the saturated state. This also suggests that \textcolor{black}{for} unstable shear flow in systems without radiative damping $|\beta_2| \approx |\beta_1|$ \textcolor{black}{is expected}, consistent with Ref.~\cite{Paper1}. The $k_y$ dependence of the ratio $|\beta_1/\beta_2|$ roughly follows that of $\gamma_1$\textcolor{black}{, except that it approaches $1$, rather than $0$, at $k_y = 0$ and $k_y = \kxeq$}.

\begin{figure}
	\includegraphics[width=8cm]{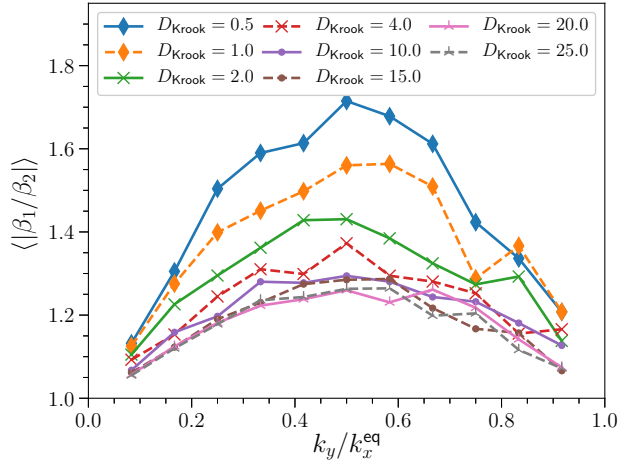}
	\caption{Time-averaged ratio of unstable to stable mode amplitude in saturation at each $k_y$ for a range of $\Dk$. Between $\Dk = 0.5$ and $4$, increasing $\Dk$ \textcolor{black}{generally} push\textcolor{black}{es} the ratio of amplitudes closer to unity. Above $\Dk = 4$, increasing $\Dk$ has a less pronounced impact on the ratio. We stress that, in both regimes, the mean flow amplitude and Reynolds stress increase with $\Dk$.}
	\label{Dk scan}
\end{figure}

Figure \ref{Dk scan} shows \textcolor{black}{how the time-averaged large-scale values of $|\beta_2/\beta_2|$ in saturation vary with $\Dk$}. The shape of the curves remains fairly consistent as $\Dk$ changes. Two regimes are apparent: below $\Dk = 4$, increasing $\Dk$ drives the ratio $|\beta_1/\beta_2|$ closer to unity, while above $\Dk = 4$ the ratio is significantly less affected. \textcolor{black}{This behavior is consistent with the notion that reinforcement of the unstable profile by larger $\Dk$ allows $\beta_2$ to be nonlinearly pumped to its maximal level, whereas for smaller $\Dk$ the quasilinear depletion of the profile cuts off the pumping of $\beta_2$ before it reaches its maximal level.} Note that $f_2$ tends to reduce \textcolor{black}{the Reynolds stress} $\tau$, suggesting that the increase in $\tau$ with $\Dk$ must be due to an increase in overall fluctuation level, rather than a change in $|\beta_1/\beta_2|$.

\subsection{\textcolor{black}{Influence of stable modes in} analytical models}
To better understand how \textcolor{black}{the unstable and stable modes} affect the mean flow in saturation, we develop a reduced model that \textcolor{black}{expresses} the mean flow amplitude in terms of $\beta_1$ and $\beta_2$. The model considers a 2D inviscid, incompressible fluid, assumes $\phi \approx \beta_1 \phi_1 + \beta_2 \phi_2$ \textcolor{black}{for $0 < k_y/\kxeq < 1$}, and \textcolor{black}{assumes} that the force applied by the Krook operator balances the turbulent Reynolds stress in saturation. These assumptions are consistent with the findings presented in Figs.~\ref{force balance} and \ref{reduced phi contours}.

For perturbations about a sinusoidal equilibrium flow, the linearized system becomes Eq.~\eqref{Fourier Rayleigh's}, which was derived in Sec.~II but is repeated here:
\begin{equation*}
\omega_j(k_x^2+k_y^2)\tilde{\phi}_j + \frac{k_y V_0}{2}\left[(k_x^2-2k_x\kxeq+k_y^2)\tilde{\phi}_j^{-}+(k_x^2+2k_x\kxeq+k_y^2)\tilde{\phi}_j^{+}\right]=0. 
\end{equation*}
Equation \eqref{Fourier Rayleigh's} can be \textcolor{black}{expressed as} a matrix equation $\omega_j \vec{\phi}_j + \mathbf{M} \vec{\phi}_j = 0$ where the \textcolor{black}{components} of $\vec{\phi}_j$ are $\tilde{\phi}_j$ at different $k_x$ and the dimension of \textcolor{black}{$\tilde{\phi}_j$ and $\mathbf{M}$} is infinite. Reasonable approximations \textcolor{black}{of} the eigenmodes and eigenvalues can be obtained by solving the matrix equation \textcolor{black}{with} $\tilde{\phi}_j \textcolor{black}{\neq 0}$ for some finite number of $k_x$ values, and $\tilde{\phi}_j = 0$ for all other $k_x$. This has previously been found useful in similar KH and tearing mode calculations \cite{Pessah}. For example, solving the system with $k_x = 0, \pm \kxeq$ yields
\begin{align*}
\omega_1 &= -\frac{ik_yV_0}{\kappa} &  \vec{\phi}_1 &= (1,-i \kappa,1)^T\\
\omega_2 &= \frac{ik_yV_0}{\kappa} &  \vec{\phi}_2 &= (1,i \kappa,1)^T\\
\omega_3 &= 0 & \vec{\phi}_3 &= (-1,0,1)^T,
\end{align*}
where the vectors are written in the form $(\tilde{\phi}(-\kxeq), \tilde{\phi}(0), \tilde{\phi}(\kxeq))^T$ and $\kappa(k_y) \equiv \sqrt{2((\kxeq)^2+k_y^2)/((\kxeq)^2-k_y^2)}$.

To arrive at an expression of force balance between the Reynolds stress and Krook drive, we return to Eq.~\eqref{mean nonlinear vorticity}, which we repeat here: 
\begin{equation*}
\frac{\partial}{\partial t}\tilde{\phi}(\kxeq,0) + \sum_{\mathbf{k}'} \frac{k'_y}{\kxeq}((\kxeq - k'_x)^2 + k'^2_y)\tilde{\phi}(k'_x, k'_y)\tilde{\phi}(\kxeq - k'_x, -k'_y) = - \Dk \tilde{\phi}(\kxeq,0). 
\end{equation*}
Considering a steady state where $\partial \tilde{\phi}(\kxeq, 0) /\partial t = 0$, and assuming $\tilde{\phi} = \beta_1 \tilde{\phi}_1 + \beta_2 \tilde{\phi}_2$ with just the $k_x = 0, \pm \kxeq$ Fourier modes considered, Eq.~\eqref{mean nonlinear vorticity} can be manipulated to yield
\begin{equation}\label{reduced model}
\tilde{\phi}(\kxeq, 0) = \frac{2 i \kxeq}{\Dk} \sum_{k_y' > 0} k_y' \kappa(k_y') \left( |\beta_1|^2 - |\beta_2|^2 \right).
\end{equation}

\begin{figure}
	\includegraphics[width=16cm]{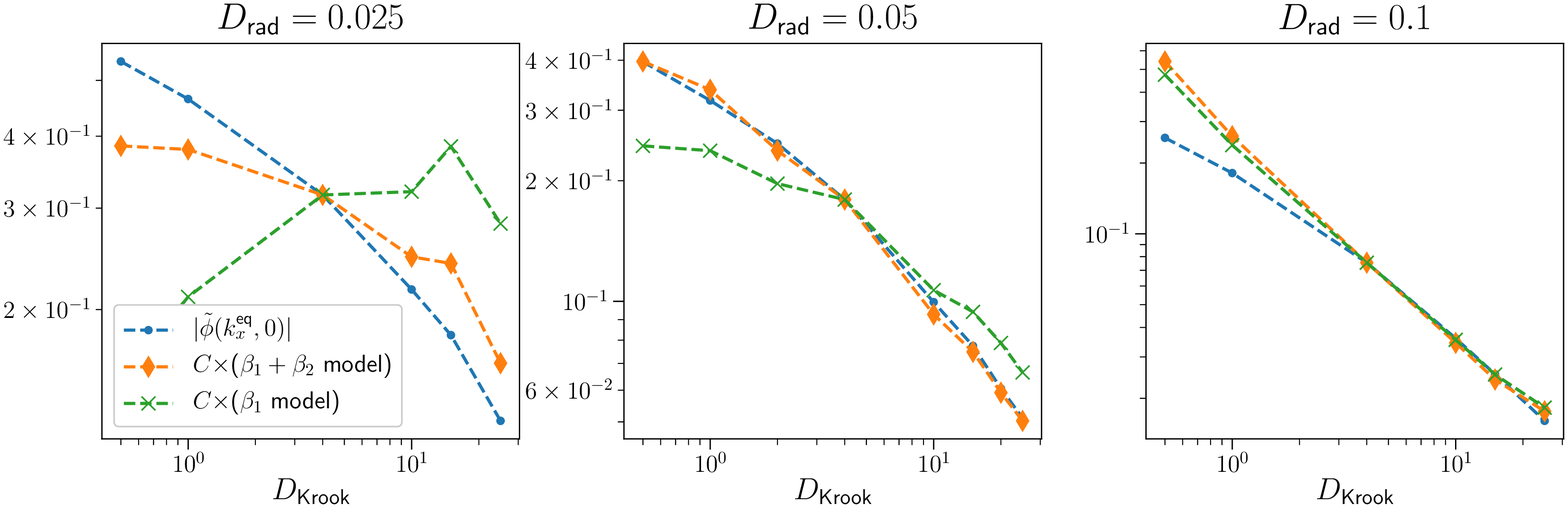}
	\caption{A comparison between the true value of $|\phi(\kxeq, 0)|$ (blue dots) and values predicted by Eq.~\eqref{reduced model} with (orange diamonds) and without (green crosses) the stable mode contribution $\beta_2$ included over multiple values of $\Dk$ and $\Drad$. In each frame, the models are scaled by a constant coefficient to match the true value at $\Dk = 4$ so that the scaling with $\Dk$ can be investigated, rather than the absolute agreement. For the base case $\Drad = 0.05$ (center frame), the scaling of $|\phi(\kxeq, 0)|$ with $\Dk$ is qualitatively captured by the model with $\beta_2$ neglected, however the scaling is significantly improved when stable modes are included. For the larger value of $\Drad$ (right frame), where stable modes are \textcolor{black}{largely} suppressed in saturation (c.f.~Fig.~\ref{rad scan}), including $\beta_2$ in the model \textcolor{black}{produces little change}, and decent quantitative agreement is observed by the model both with and without $\beta_2$ included. For the smaller value of $\Drad$ (left frame), where stable modes are more important, the model fails to even qualitatively agree with simulations unless $\beta_2$ is included\textcolor{black}{, in which case the overall trend is captured}.}
	\label{2d scan}
\end{figure}

From here, values for $|\beta_1|$ and $|\beta_2|$ can be inserted to arrive at values for the mean flow amplitude. In other systems, $|\beta_j|$ have been calculated using statistical closures \cite{Baver2002,TerryClosure,Hegna}. \textcolor{black}{Extending the above approach with s}uch a calculation would yield a complete model, but is outside the scope of this paper. Instead, we insert values of $|\beta_1|$ and $|\beta_2|$ from \textcolor{black}{nonlinear} simulations into Eq.~\eqref{reduced model}. Our interest is in the scaling of $\tilde{\phi}(\kxeq, 0)$ with $\Dk$, and what role $\beta_2$ plays in that scaling. \textcolor{black}{Thus}, we perform simulations with a range of $\Dk$ and compare three quantities: the time-averaged value of $\tilde{\phi}(\kxeq, 0)$ in saturation, the result of Eq.~\eqref{reduced model} using both $\beta_1$ and $\beta_2$, and the result obtained when $\beta_2 = 0$ is assumed. In Fig.~\ref{2d scan}, this comparison is made for three values of $\Drad$, corresponding to three systems with varying degrees of stable mode excitation (recall stable modes are more excited at lower values of $\Drad$, see Fig.~\ref{rad scan}). For $\Drad = 0.05$, where stable modes were shown in Fig.~\ref{base case betas} to be significant, the $\beta_2 = 0$ model is qualitatively correct, but significantly improved when $\beta_2$ is properly included. For $\Drad = 0.025$, where stable modes are even more important, the $\beta_2 = 0$ model fails to even capture the decrease in $|\tilde{\phi}(\kxeq, 0)|$ with $\Dk$, while the model that includes $\beta_2$ does capture the correct qualitative \textcolor{black}{and often even quantitative} behavior. At $\Drad = 0.1$, where stable modes are significantly \textcolor{black}{weakened}, their inclusion does not have a significant impact on the model. For each value of $\Drad$, the two models are scaled by a constant so that they agree with the simulation results at $\Dk = 4$, which is where the change in scaling with respect to $\Dk$ was noted in Fig.~\ref{Dk scan}. \textcolor{black}{(}It is the scaling properties of the models that we are assessing, not the absolute values.\textcolor{black}{)} Note that this model neglects all eigenmodes except $\phi_1$ and $\phi_2$, including the modes with nonzero $k_x$ at noninteger multiples of $\kxeq$. 

Comparing the two models at different values of $\Drad$ demonstrates that when stable modes are excited in this system as in Fig.~\ref{base case betas}, they not only modify the shape of the flow, as shown in Fig.~\ref{reduced phi contours}, but have an important impact on how the system responds to forcing. By nonlinearly transferring energy into large-scale stable modes, the fluctuating flow adjusts in a way that changes the feedback onto the large-scale mean flow, thus affecting how the system is forced. We also note that Eq.~\eqref{reduced model} was derived assuming an inviscid fluid, while the inserted values for $|\beta_1|$ and $|\beta_2|$ were obtained from gyrokinetic simulations with finite $\Dhyp$, suggesting gyrokinetic effects \textcolor{black}{may} not play a significant role in the eigenmode excitations and force balance in this system.

\section{Conclusions}
\textcolor{black}{We have studied an unstable gyrokinetic shear flow, finding that the system includes a conjugate stable eigenmode that is nonlinearly driven to a large amplitude leading into saturation, and continues to make important contributions to the Reynolds stress in the quasi-stationary turbulent state, except at high values of radiative damping. This demonstrates that previous findings on the role of stable eigenmodes in shear-flow instability saturation of a fluid shear layer are consistent with the quasi-stationary turbulent state of a gyrokinetic periodic shear flow. Furthermore, our results point to the potential for reduced models of shear-driven turbulence to be significantly improved by including stable mode physics.}

We have investigated the \textcolor{black}{saturation} of a linearly unstable $E \times B$ shear flow in gyrokinetics \textcolor{black}{as it relates to the full eigenmode spectrum. We find that the gyrokinetic system} compares well with \textcolor{black}{its} hydrodynamic \textcolor{black}{counterpart} with regards to the unstable mode, \textcolor{black}{as well as} the rest of the spectrum. Specifically, the dissipationless linear operator includes a single conjugate stable eigenmode for every unstable eigenmode, along with a continuum of marginally stable modes. \textcolor{black}{N}onlinear simulations \textcolor{black}{characterize the} behavior \textcolor{black}{of the flow} in saturation, \textcolor{black}{and we examine cases} both with and without an external driving term. \textcolor{black}{The} drive is implemented in the form of a Krook operator, and reinforces the unstable mean flow in a manner similar to Kolmogorov flow.

In simulations without the drive term, the system lacks any energy injection to \textcolor{black}{maintain} the unstable equilibrium\textcolor{black}{. This causes} fluctuations to quickly relax the unstable flow shear once nonlinear interactions become significant, and the \textcolor{black}{turbulence subsequently} decays. In simulations with forcing, we include a scale-independent radiative damping term that prevents accumulation of energy at the largest scales, and allows a quasi-stationary state of driven turbulence. In driven simulations, a partial relaxation of the mean flow is still observed, with the final state mostly determined by a force balance between the Krook drive and the turbulent Reynolds stress.

With a well-resolved system of quasi-stationary, driven turbulence, we investigate the role of different linear eigenmodes by performing an eigenvalue decomposition, where the turbulent state is expressed as a linear combination of the eigenmodes. The evolution of the dominant pair of stable and unstable modes leading into saturation compares well with previous analytic calculations of an inviscid fluid shear layer \cite{Paper1}, and the ensuing excitation of the stable mode in the turbulent state is \textcolor{black}{broadly} consistent with previous findings in plasma microturbulence \cite{Makwana}. By demonstrating that the role of stable modes in shear-flow instability saturation is consistent with their role in the fully-developed turbulent system, we have extended the set of systems in which instability saturation analyses has proven to be predictive of the turbulent state to include fully global fluid instabilities, further motivating these sorts of analyses in other global instabilities where stable modes exist, such as the magnetorotational instability \cite{Clark}.

The significant excitation of linearly stable modes in the saturated state indicates that an important aspect of shear-driven turbulence is this previously-neglected tendency for large-scale fluctuations to lose energy back into the mean flow via the linear operator. This idea is in contrast with the standard picture of instability-driven turbulence, where it is assumed that the largest scales are dominated by a balance between linear energy injection and nonlinearly energy transfer to \textcolor{black}{smaller} scales. While many other modes are also excited in the saturated state, we have shown that the stable/unstable pair of modes \textcolor{black}{is sufficient to capture many aspects of the flow}. This also presents a significant modification to the existing understanding of shear-driven turbulence, where reduced models generally assume that large-scale fluctuations are dominated by unstable modes alone \cite{Gaster,Liou,Nikitopoulos,Horton}. 


Consistent with previous work where the conjugate symmetry between unstable/stable pairs of modes was broken with dissipative terms \cite{Terry2009}, we find that the added radiative damping term, which increases the damping rate of the stable mode and reduces the growth rate of the unstable mode, suppresses the importance of the stable mode relative to the unstable one. This is observed by comparing the amplitudes of the two modes for a range of radiative damping values. Making use of the observations that the gyrokinetic and fluid systems behave similarly, that the mean flow amplitude at saturation is determined by force balance between driving and Reynolds stress, and that the stable and unstable modes alone describe large-scale fluctuations well, we construct a reduced model that allows us to examine the role of stable modes in determining the mean flow in saturation\textcolor{black}{. The model results} in an equation where the contributions from stable modes can be isolated from unstable modes. We find that lower values of radiative damping, where stable modes \textcolor{black}{exhibit} higher amplitude\textcolor{black}{s}, require the inclusion of stable modes in the model in order for it to be even qualitatively correct. At higher radiative damping\textcolor{black}{,} where stable modes are suppressed, their inclusion in the model has no significant impact on its performance. Thus, in shear-flow systems where stable modes play an important role in instability saturation, they may also be expected to play an important role in understanding how fluctuations affect the mean flow, and thus how the system responds to external forcing. \textcolor{black}{We further conclude that when effects observed to change turbulence characteristics also break the conjugate symmetry of an unstable/stable eigenmode pair \cite{Palotti}, the change in turbulence may be related to differences in stable mode excitation.}

\begin{acknowledgments}
The authors would like to thank B.N.~Rogers, W.~Dorland, and M.E.~Pessah for valuable discussions and insights. Partial support for this work was provided by the National Science Foundation under Award PHY-1707236, the Wisconsin Alumni Research Foundation, the Vilas Trust, and the U S Department of Energy, Office of Science, Fusion Energy Sciences, under Award No.~DE-FG02-89ER53291. Computing resources were provided by the National Science Foundation through XSEDE computing resources, allocation No.~TG-PHY130027.
\end{acknowledgments}


\begin{thebibliography}{9}
\bibitem{Balbus}
S.A.~Balbus and J.F.~Hawley, Rev.~Mod.~Phys.~\textbf{70}, 1 (1998).
\bibitem{Vanon}
R.~Vanon and G.I.~Ogilvie, Mon.~Not.~Roy.~Ast.~Soc.~\textbf{463}, 3725 (2016).
\bibitem{Faganello}
M.~Faganello and F.~Califano, J.~Plasma Phys.~\textbf{83}, 535830601 (2017).
\bibitem{Rogers2000}
B.N.~Rogers, W.~Dorland, and M.~Kotschenreuther, Phys.~Rev.~Lett.~\textbf{85}, 25 (2000).
\bibitem{Chandrasekhar}
S.~Chandrasekhar, \textit{Hydrodynamic and Hydromagnetic Stability} (Oxford University Press, 1961).
\bibitem{DrazinReid}
P.G.~Drazin and W.H.~Reid, \textit{Hydrodynamic Stability} (Cambridge University Press, 1981).
\bibitem{Gaster}
M.~Gaster, E.~Kit, and I.~Wygnanski, J.~Fluid Mech.~\textbf{150}, 23 (1985).
\bibitem{Palotti}
M.L.~Palotti, F.~Heitsch, E.G.~Zweibel, and Y.-M.~Huang, Astrophys.~J.~\textbf{678}, 234 (2008).
\bibitem{Liou}
W.W.~Liou and P.J.~Morris, Phys.~Fluids A \textbf{4}, 2798 (1992).
\bibitem{Nikitopoulos}
D.E.~Nikitopoulos and J.T.C~Liu, Phys.~Fluids \textbf{13}, 966 (2001).
\bibitem{Horton}
W.~Horton, T.~Tajima, and T.~Kamimura, Phys.~Fluids \textbf{30}, 3485 (1987).
\bibitem{Henri}
P.~Henri, S.S.~Cerri, F.~Califano, F.~Pegoraro, C.~Rossi, M.~Faganello, O.~\v{S}ebek, P.M.~Tr\'{a}vn\'{i}\v{c}ek, P.~Hellinger, J.T.~Frederiksen, A.~Nordlund, S.~Markidis, R.~Keppens, and G.~Lapenta, Phys.~Plasmas \textbf{20}, 102118 (2013).
\bibitem{Paper1}
A.E.~Fraser, P.W.~Terry, E.G.~Zweibel, and M.J.~Pueschel, Phys.~Plasmas \textbf{24}, 062304 (2017).
\bibitem{Terry2018}
P.W.~Terry, B.J.~Faber, C.C.~Hegna, V.V.~Mirnov, M.J.~Pueschel, and G.G.~Whelan, Phys.~Plasmas \textbf{25}, 012308 (2018).
\bibitem{Hegna}
C.C.~Hegna, P.W.~Terry, and B.J.~Faber, Phys.~Plasmas \textbf{25}, 022511 (2018).
\bibitem{Whelan}
G.G.~Whelan, M.J.~Pueschel, and P.W.~Terry, Phys.~Rev.~Lett.~\textbf{120}, 175002 (2018).
\bibitem{Jenko}
F.~Jenko, W.~Dorland, M.~Kotschenreuther, and B.N.~Rogers, Phys.~Plasmas \textbf{7}, 1904 (2000).
\bibitem{genecode}
See http://www.genecode.org for code details and access.
\bibitem{HatchLeft}
D.R.~Hatch, P.W.~Terry, F.~Jenko, F.~Merz, M.J.~Pueschel, W.M.~Nevins, and E.~Wang, Phys.~Plasmas \textbf{18}, 055706 (2011).
\bibitem{TerryLeft}
P.W.~Terry, K.D.~Makwana, M.J.~Pueschel, D.R.~Hatch, F.~Jenko, and F.~Merz, Phys.~Plasmas \textbf{21}, 122303 (2014).
\bibitem{Rogers2005}
B.N.~Rogers and W.~Dorland, Phys.~Plasmas \textbf{12}, 062511 (2005).
\bibitem{Platt}
N.~Platt, L.~Sirovich, and N.~Fitzmaurice, Phys.~Fluids A \textbf{3}, 681 (1991).
\bibitem{Musacchio}
S.~Musacchio and G.~Boffetta, Phys.~Rev.~E \textbf{89}, 023004 (2014).
\bibitem{Lucas}
D.~Lucas and R.~Kerswell, J.~Fluid Mech.~\textbf{750}, 518 (2014).
\bibitem{Goodman}
J.~Goodman and G.~Xu, Astrophys.~J.~\textbf{432}, 213 (1994).
\bibitem{Latter2009}
H.N.~Latter, P.~Lesaffre, and S.A.~Balbus, Mon.~Not.~Roy.~Ast.~Soc.~\textbf{394}, 715 (2009).
\bibitem{PessahGoodman}
M.E.~Pessah and J.~Goodman, Astrophys.~J.~\textbf{698}, L72 (2009).
\bibitem{Pessah}
M.E.~Pessah, Astrophys.~J.~\textbf{716}, 1012 (2010).
\bibitem{Latter2010}
H.N.~Latter, S.~Fromang, and O.~Gressel, Mon.~Not.~Roy.~Ast.~Soc.~\textbf{406}, 848 (2010).
\bibitem{Longaretti}
P.Y.~Longaretti and G.~Lesur, Astron.~Astrophys.~\textbf{516}, 51 (2010).
\bibitem{Squire}
J.~Squire, E.~Quataert, and M.W.~Kunz, J.~Plasma Phys.~\textbf{83}, 905830613 (2017).
\bibitem{Kim}
J.-H.~Kim and P.W.~Terry, Phys.~Plasmas \textbf{17}, 112306 (2010).
\bibitem{PueschelKrook}
M.J.~Pueschel, D.~Told, P.W.~Terry, F.~Jenko, E.G.~Zweibel, V.~Zhdankin, and H.~Lesch, Astrophys.~J.~Suppl.~Ser.~\textbf{213}, 30 (2014).
\bibitem{MarstonKrook}
J.B.~Marston, E.~Conover, and T.~Schneider, J.~Atmos.~Sci.~\textbf{65}, 1955 (2008).
\bibitem{Terry2009}
P.W.~Terry, D.A.~Baver, and D.R.~Hatch, Phys.~Plasmas \textbf{16}, 122305 (2009).
\bibitem{Brizard}
A.J.~Brizard and T.S.~Hahm, Rev.~Mod.~Phys.~\textbf{79}, 421 (2007).
\bibitem{Merz}
F.~Merz, Ph.D.~thesis, University of M{\"u}nster, 2009.
\bibitem{Pueschel2011}
M.J.~Pueschel, F.~Jenko, D.~Told, and J.~B{\"u}chner, Phys.~Plasmas \textbf{18}, 112102 (2011).
\bibitem{Pueschel2013a}
M.J.~Pueschel, D.R.~Hatch, T.~G{\"o}rler, W.M.~Nevins, F.~Jenko, P.W.~Terry, and D.~Told, Phys.~Plasmas \textbf{20}, 102301 (2013).
\bibitem{DedalusLib}
K.J.~Burns, G.M.~Vasil, J.S.~Oishi, D.~Lecoanet, and B.~Brown, Astrophys.~Source Code Lib.~record ascl:1603.015.
\bibitem{DedalusWeb}
See http://dedalus-project.org for details.
\bibitem{MJCPC}
M.J.~Pueschel, T.~Dannert, and F.~Jenko, Comput.~Phys.~Commun.~\textbf{181}(8), 1428-1437 (2010).
\bibitem{TobiasFriction}
S.M.~Tobias, K.~Dagon, and J.B.~Marston, Astrophys.~J.~\textbf{727}, 12 (2011).
\bibitem{Reynolds-Barredo}
J.M.~Reynolds-Barredo, D.E.~Newman, P.W.~Terry, and R.~Sanchez, Europhys.~Lett. \textbf{155}, 34002 (2016).
\bibitem{Waleffe}
F.~Waleffe, Phys.~Fluids \textbf{7}, 3060 (1995).
\bibitem{GENE-EV}
M.~Kammerer, F.~Merz, and F.~Jenko, Phys.~Plasmas \textbf{15}, 052102 (2008).
\bibitem{Case}
K.M.~Case, Phys.~Fluids \textbf{3}, 143 (1960).
\bibitem{Brandstater}
A.~Brandst{\"a}ter, J.~Swift, H.L.~Swinney, A.~Wolf, J.D.~Farmer, E.~Jen, and P.J.~Crutchfield, Phys.~Rev.~Lett.~\textbf{51}, 1814 (1983).
\bibitem{Terry2006}
P.W.~Terry, D.A.~Baver, and S.Gupta, Phys.~Plasmas \textbf{13}, 022307 (2006).
\bibitem{Makwana}
K.D.~Makwana, P.W.~Terry, J.-H.~Kim, and D.R.~Hatch, Phys.~Plasmas \textbf{18}, 012302 (2011).
\bibitem{MJBen}
M.J.~Pueschel, B.J.~Faber, J.~Citrin, C.C.~Hegna, P.W.~Terry, and D.R.~Hatch, Phys.~Rev.~Lett.~\textbf{116}, 085001 (2016).
\bibitem{Ben2018}
B.J.~Faber, M.J.~Pueschel, P.W.~Terry, C.C.~Hegna, and J.E.~Roman, Stellarator microinstabilities and turbulence at low magnetic shear, submitted to J.~Plasma Phys.~(2018).
\bibitem{HatchPseudo}
D.R.~Hatch, F.~Jenko, A.~Ba{\~n}{\'o}n Navarro, V.~Bratanov, P.W.~Terry, and M.J.~Pueschel, New J.~Phys.~\textbf{18}, 075018 (2016)
\bibitem{Baver2002}
D.A.~Baver, P.W.~Terry, R.~Gatto, and E.~Fernandez, Phys.~Plasmas \textbf{9}, 3318 (2002).
\bibitem{TerryClosure}
P.W.~Terry and R.~Gatto, Phys.~Plasmas \textbf{13}, 062309 (2006).
\bibitem{Clark}
S.E.~Clark, Ph.D.~thesis, Columbia University, 2017.
\end{thebibliography}
\end{document}